%% file: TAPAS-7-final.tex
\documentclass{aa}  

\usepackage{graphicx}
\usepackage{txfonts}
\usepackage{siunitx}
\usepackage{multirow}
\usepackage{lscape}
\usepackage{color}
\usepackage{longtable}
\def\Mjup{\hbox{$\thinspace M_{\mathrm{J}}$}}
\def\Msun{\hbox{$\thinspace M_{\odot}$}}
\def\Rsun{\hbox{$\thinspace R_{\odot}$}}
\def\Lsun{\hbox{$\thinspace L_{\odot}$}}
\def\Teff{\hbox{$\thinspace T_{\mathrm{eff}}$}}

\def\kms{\hbox{$\thinspace {\mathrm{km~s^{-1}}}$}}
\def\ms{\hbox{$\thinspace {\mathrm{m~s^{-1}}}$}}

\def\au{\hbox{$\thinspace \mathrm{au}$}}
\def\sjit{\hbox{$\sigma_{\mathrm{jitter}}$}}
\def\srv{\hbox{$\sigma_{\mathrm{RV}}$}}
\def\sbs{\hbox{$\sigma_{\mathrm{BIS}}$}}

\def\Prot{\hbox{$P_{\mathrm{rot}}$}}

\def\llsun{\hbox{$\thinspace \log L/L_{\odot}$}} 
 
\def\shk{\hbox{$S_{\mathrm{HK}}$}}
\def\ha{\hbox{$I_{\mathrm{H_{\alpha}}}$}}

\def\starA{HD~4760\thinspace} 
\def\starB{BD+02 3313\thinspace} 
\def\starC{TYC 0434-04538-1\thinspace} 
\def\starE{HD~96992\thinspace} 

\begin{document}

  \title{Tracking Advanced Planetary Systems (TAPAS) with HARPS-N.  
  \thanks{Based on observations obtained with the Hobby-Eberly Telescope, which is a joint project 
  of the University of Texas at Austin, the Pennsylvania State University, 
Stanford University, Ludwig-Maximilians-Universit\"at M\"unchen, and Georg-August-Universit\"at G\"ottingen.}
\thanks{Based on observations made with the Italian Telescopio Nazionale Galileo (TNG) operated on the island of
 La Palma by the Fundaci\'on Galileo Galilei of the INAF (Istituto Nazionale di Astrofisica) 
 at the Spanish Observatorio del Roque de los Muchachos of the Instituto de Astrof\'{\i}sica de Canarias.}
}

   \subtitle{VII.  Elder suns with low-mass companions  }

   \titlerunning{TAPAS VII.  Elder suns with low-mass companions }
   \authorrunning{A. Niedzielski et al.}

 \author{
         A. Niedzielski
          \inst{1}  
\and
         E. Villaver
         \inst{2,3}
 \and                         
        M. Adam\'ow
         \inst{4,5}
\and                         
          K. Kowalik
          \inst{4}   
          \and       
         A. Wolszczan
          \inst{6,7}   
\and                          
         G. Maciejewski
          \inst{1}        
                  }
                   
\institute{
            Institute of Astronomy, Faculty of Physics, Astronomy and Applied Informatics, Nicolaus Copernicus University in Toru\'n, Gagarina 11, 87-100 Toru\'n, Poland,  
            \email{Andrzej.Niedzielski@umk.pl}
\and
            Departamento de F\'{\i}sica Te\'orica, Universidad Aut\'onoma de Madrid, Cantoblanco 28049 Madrid, Spain, \email{eva.villaver@uam.es}   
\and
	Centro de Astrobiología (CAB, CSIC-INTA), ESAC Campus Camino Bajo del Castillo, s/n, Villanueva de la Cañada, E-28692 Madrid, Spain
\and
            National Center for Supercomputing Applications, University of Illinois, Urbana-Champaign, 1205 W Clark St, MC-257, Urbana, IL 61801, USA   
\and
           Center for Astrophysical Surveys, National Center for Supercomputing Applications, Urbana, IL, 61801, USA
\and
            Department of Astronomy and Astrophysics, Pennsylvania State University, 525 Davey Laboratory, University Park, PA 16802, USA
            \email{alex@astro.psu.edu}
\and
            Center for Exoplanets and Habitable Worlds, Pennsylvania State University, 525 Davey Laboratory, University Park, PA 16802, USA
            }

   \date{Received;accepted}

 
  \abstract
   { We present the current status of and 
   new results from our search 
   for 
  exoplanets in a sample of 
   solar-mass, 
   evolved stars observed with the HARPS-N and the 3.6-m Telescopio Nazionale Galileo (TNG), and the High Resolution Spectrograph (HRS) and the 9.2-m Hobby Eberly Telescope (HET).  }
   {The aim of this project is to detect and characterise planetary-mass companions to  solar-mass stars  
   in a sample of 122 targets at various stages of evolution from the main sequence (MS) to the red giant branch (RGB), mostly sub-gaints and giants, selected from the Pennsylvania-Toruń Planet Search (PTPS) sample, and use this sample to study relations between stellar properties, such as  metallicity, luminosity, and the planet occurrence rate.  }
   {This work is based on precise radial velocity (RV) measurements. We have observed the program stars for up to 11 years with the HET/HRS and the TNG/HARPS-N.
    }
   {We present the analysis of RV measurements with the HET/HRS and the TNG/HARPS-N 
   of four solar-mass stars, HD 4760, HD 96992 , BD+02 3313, and  TYC 0434-04538-1. We found that:
      HD 4760  hosts a companion with a minimum mass of $13.9\Mjup$ ($a=1.14$~au, $e=0.23$); 
   HD 96992 is a  host to a $m\sin i=1.14\Mjup$  companion on a $a=1.24$ au and $e=0.41$ orbit, and 
TYC 0434-04538-1 hosts an $m\sin i=6.1\Mjup$ companion on a $a=0.66$ au and $e=0.08$ orbit. In the case of BD+02 3313 we found a correlation between the measured RVs and one of the stellar activity indicators, suggesting that the observed RV variations may 
originate in either stellar activity or be caused by the presence of an unresolved companion. We also discuss the current status of the project and a statistical analysis of the RV variations in our sample of target stars.

}
   { In our sample of 122 solar-mass stars,  $49\pm5\%$ of them appear to be single, and $16\pm3\%$ are spectroscopic binaries.
   The three giants hosting low-mass companions presented in this paper add to the six ones previously identified in the sample.}


   \maketitle
%

\section{Introduction}

After the discovery of the first exoplanetary system around a pulsar (PSR 1257+12 b, c, d --  \citealt{Wolszczan1992}) 
with the pulsar timing technique, and of the first exoplanet  orbiting  a solar-type star (51 Peg b -- \citealt{Mayor1995})
with the precise velocimetry, the photometric observations of planetary transits have proved  
to be the most successful way of detecting exoplanets. 

Nearly 3000 out of about 4300 exoplanets  were detected with the planetary transit method,
most of them by just one project, Kepler/K2 \citep{2010Sci...327..977B}.
Detailed characterisation of these systems requires both photometric (transits) and spectroscopic
(radial velocities, abundances) observations, but not all of them are available for spectroscopic follow-up with ground-based instruments, due to the faintness of the hosts. This emphasizes the need
for missions such as TESS \citep{TESS} and PLATO \citep{PLATO}.

Our knowledge of exoplanets orbiting the solar-type or less massive stars on the MS is quite extensive due to combined output of the RV
and transit searches (see \citealt{2015ARA&A..53..409W} for a review). The domain 
of larger orbital separations or more evolved hosts clearly requires more exploration.

 \begin{table*}
\centering
\tiny
\caption{Basic parameters of the program stars. }
  \begin{tabular}{lllllllll}
\hline
   Star           & $\Teff$[K]  & $\log g$      & $[$Fe/H$]$     & $\log L/\Lsun$             & $M/\Msun$                 & $R/\Rsun$     & $v\sin i\;[\!\kms]$ & {$\Prot\;[\mathrm{days}]$}  \\
\hline \hline
HD 4760 			& 4076$\pm$15 & 1.62$\pm$0.08 & -0.91$\pm$0.09 & {2.93$\pm$0.11} & 1.05$\pm$0.19        & 42.4$\pm$9.2  & $1.40\pm1.10$ & $1531 \pm 1535$   \\ 
HD 96992			& 4725$\pm$10 & 2.76$\pm$0.04 & -0.45$\pm$0.08 & 1.47$\pm$0.09              & 0.96$\pm$0.09             & 7.43$\pm$1.1  & $1.90\pm0.60$ & $198\pm 92$ \\  
BD+02 3313 		& 4425$\pm$13 & 2.64$\pm$0.05 &  0.10$\pm$0.07 & 1.44$\pm$0.24              & 1.03$\pm$0.03             &  8.47$\pm$1.53& $1.80\pm0.60$& $238\pm122$ \\  
TYC 0434-04538-1	& 4679$\pm$10 & 2.49$\pm$0.04 & -0.38$\pm$0.06 & 1.67$\pm$0.09              & 1.04$\pm$0.15             & 9.99$\pm$1.6  & $3.00\pm0.40$& $169 \pm 49 $ \\  

\hline
\end{tabular}
\label{parameters}
\end{table*}


So far, the RV searches for exoplanets orbiting more evolved stars, like
Lick K-giant Survey \citep{Frink2002}, 
Okayama Planet Search\citep{Sato2003}, 
Tautenberg Planet Search \citep{Hatzes2005},  
Retired A Stars and Their Companions \citep{Johnson2007},  
PennState - Toru\'n  Planet Search \citep{Niedzielski2007, Niedzielski2008, NiedzielskiWolszczan2008} 
or Boyunsen Planet Search \citep{Lee2011}, 
have resulted in  a rather modest population of 112 substellar companions in 102 systems\footnote{\tiny{https://www.lsw.uni-heidelberg.de/users/sreffert/giantplanets/giantplanets.php}}.

The Pennsylvania-Toru\'n Planet Search (PTPS) is one of the most extensive RV searches for exoplanets around the evolved stars.
The project was designed to use the Hobby-Eberly Telescope \citep{Tull1998} (HET) and its High Resolution Spectrograph \citep{Ramsey1998} (HRS).
It has surveyed a sample of stars 
distributed across the northern sky, with the typical, apparent V-magnitudes between 7.5 and 10.5 mag, and the B-V colour indices  between 0.6 and 1.3.  
On the Hertzsprung-Russell (H-R) diagram, these stars occupy  an area delimited by the MS, the instability strip, and the coronal dividing line \citep{1979ApJ...229L..27L}. 
{ In total, the program sample of 885 stars contains 515 giants, 238 subgiants, and 132 dwarfs \citep{Deka-Szymankiewicz2018}.}
A detailed description of this sample, 
 including their atmospheric  
and integrated parameters (masses, luminosities, and radii), is presented  in a series of  the following papers:
\cite{Zielinski2012, Adamow2014, Niedzielski2016a, Adamczyk2016, Deka-Szymankiewicz2018}. { The first detection of a gas giant orbiting a red giant star by the  PTPS} project has been published by \cite{Niedzielski2007}. 

So far, twenty-two planetary systems have been detected by the PTPS and TAPAS projects. The most interesting ones include: 
a multiple planetary system around TYC 1422-00614-1, an evolved solar-mass, K2 giant, with two planets orbiting it \citep{TAPAS1};
the most massive, $1.9 \Msun$, red giant star TYC 3667-1280-1, hosting a warm Jupiter \citep{TAPAS4},
 and BD+48 740, a Li overabundant giant star with a planet, which possibly represents a  case of recent engulfment \citep{Adamow2012}.
Of specific interest is BD+14 4559 b, a $1.5 \Mjup$  gas giant orbiting  a $0.9 \Msun$ dwarf in an eccentric  orbit (e=0.29) at a distance of a=0.78 au from the host star \citep{Niedzielski2009b}. 
The  International Astronomical Union chose this planet and its host  on the occasion of its 100th anniversary, 
to be named by the Polish national poll organized by the  ExoWorlds project. They have been assigned  the names of Pirx and Solaris to honor the  famous Polish science fiction writer Stanisław Lem.

{ The PTPS sample is large enough to investigate planet occurrence as a function of a well-defined set of stellar parameters.}
For instance, the sample of 15 Li-rich giants has been studied in a series of papers \citep{Adamow2012, Adamow2014, Adamow2015, 2018A&A...613A..47A} and resulted in a discovery 
of 3 Li-rich giants with planetary-mass companions: BD+48 740, HD 238914, and TYC 3318-01333-1 
 and two planetary-mass companions candidates: TYC 3663-01966-1 and TYC 3105-00152-1. 
Another  interesting subsample of the PTPS  contains 115 stars with masses greater than $1.5 \Msun$. 
So far,  four giants with planets were detected in that sample: HD 95127, HD 216536,  BD+49 828 \citep{Niedzielski2015b}, 
and TYC 3667-1280-1 \citep{TAPAS4}  with masses as high as $1.87 \Msun$.
A $2.88 \Msun$ giant TYC 3663-01966-1, mentioned above, also belongs to this subsample.

There are more PTPS stars to be investigated in search for low-mass companions: these are 74  low metallicity ([Fe/H]$\le$-0.5) giant stars,
including BD +20 2457 \citep{Niedzielski2009b}  and BD +03 2562 \citep{2017A&A...606A..38V} - both with [Fe/H]$\le$-0.7), 
 57 high luminosity giants with $\llsun\ge2$, cf. 
BD +20 2457 \citep{Niedzielski2009b},
HD 103485 \citep{2017A&A...606A..38V} both with $\llsun \ge2.5$),
and a number of others.
All these investigations are still in progress. Here, we present the results for four of the program stars.

\section{Sample and observations \label{observations}}

\begin{figure}
   \centering
   \includegraphics[width=0.5\textwidth]{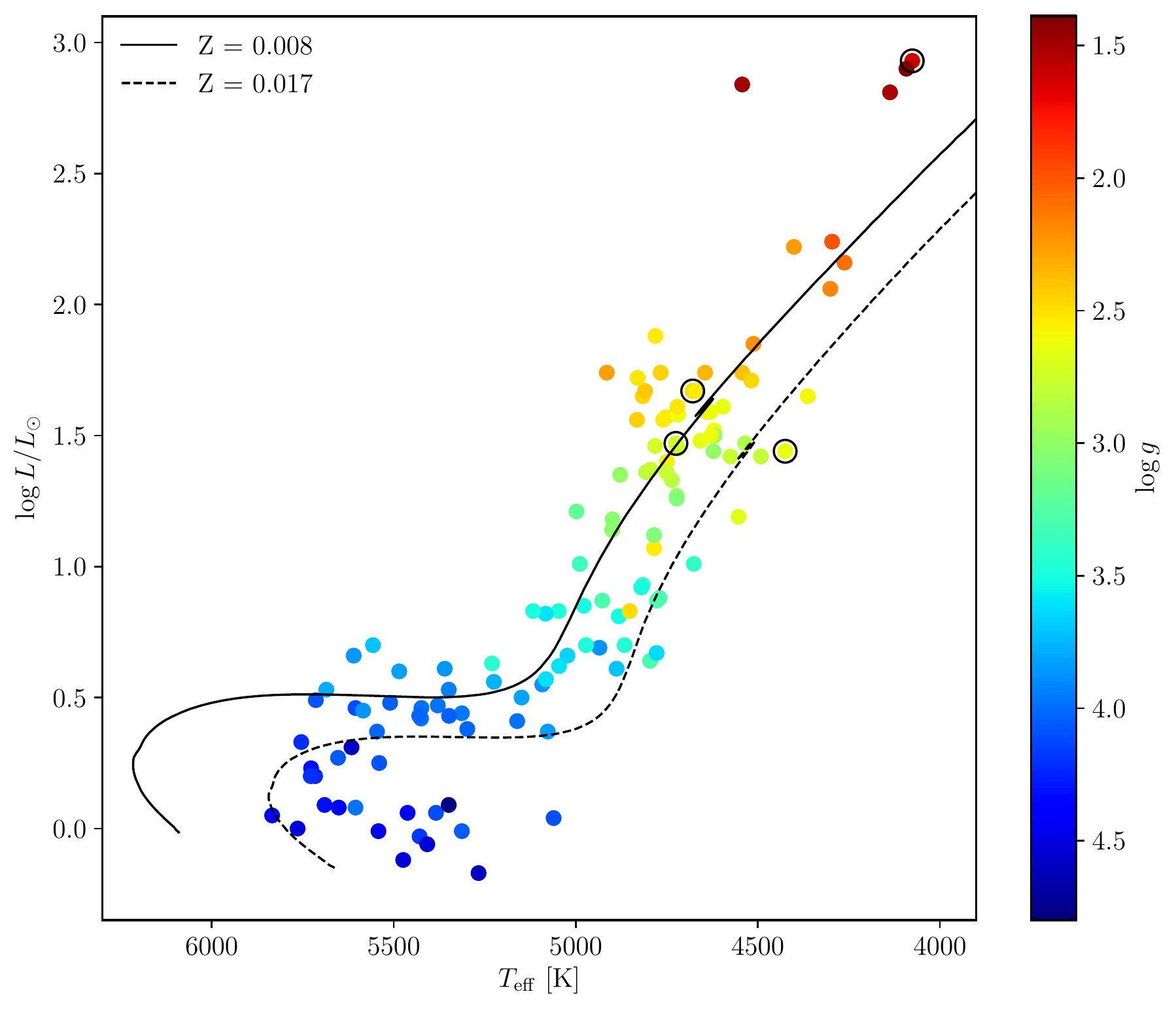}
   \caption{The Hertzsprung-Russell diagram for 122 PTPS stars with solar masses within $5\%$ uncertainty. Circles mark stars discussed in this work.}
   \label{HRD}
\end{figure}

There are 133 stars in the PTPS sample with masses in the $1\pm0.05\Msun$ range:  12 dwarfs, 39 subgiants, and 82 giants (\citealt{Deka-Szymankiewicz2018} and references therein).
Due to an insufficient RV time series coverage (less than two epochs of observations), we have removed eleven of these stars from further considerations.
Consequently, the final, complete sample of 122  solar-mass stars contains 11 dwarfs, 33 subgiants, and 78 giant stars (Fig. \ref{HRD}).
In what follows, we will call them {\it elder suns},  { representing a range of evolutionary stages (from the MS through the subgiant branch and along the RGB)  
and a range of metallicities (between [Fe/H]=-1.44 and [Fe/H]=+0.34, with [Fe/H]=-0.17 being the average).  
 However, within the estimated uncertainties, their estimated masses are all the same.
 The small group of dwarfs included in the sample represents stars similar to the Sun with different metallicities.
The sample defined this way allows us 
 to study the planet occurrence ratio as a function of stellar metallicity for a fixed solar mass.}

Here we present the results for four stars from this sample, that show RV variations appearing be caused by  low-mass companions.
Their basic atmospheric and stellar parameters are summarised  in Table \ref{parameters}.
The atmospheric parameters, $\Teff$, $\log g$,  and $[$Fe/H$]$, were derived using a strictly spectroscopic method based on the 
LTE analysis of the equivalent widths of FeI and FeII lines by \cite{Zielinski2012}. The estimates of the rotational velocities are given in \cite{Adamow2014}. The stellar parameters (masses, luminosities, and radii) were 
estimated using the Bayesian approach of  \cite{Jorgensen2005}, modified by \cite{daSilva2006} and adopted for our project by \cite{Adamczyk2016},
using the theoretical stellar models from \cite{Bressan2012}.
In the case of BD+02 3313,  we determined the luminosity  using the  \cite{Gaia2016} DR2 parallax  (see \citealt{Deka-Szymankiewicz2018} for details).

\subsection{Observations}

The spectroscopic observations presented in this paper were made  with two instruments: 
the 9.2-m  Hobby-Eberly Telescope
(HET, \citealt{Ramsey1998}) and its  High-Resolution Spectrograph (HRS,
\citealt{Tull1998}) in the queue scheduling mode \citep{Shetrone2007},
and the 3.58-meter Telescopio Nazionale Galileo (TNG) and its High
Accuracy Radial velocity Planet Searcher in the Northern hemisphere (HARPS-N,
\citealt{Cosentino2012}).
A detailed description of the adopted observing strategies and the instrumental configurations for both 
HET/HRS and TNG/HARPS-N can be found in  \cite{Niedzielski2007} and \cite{TAPAS1}.

All HET/HRS spectra were reduced with the standard IRAF\footnote{IRAF is distributed 
by the National Optical Astronomy Observatories, which are operated by the Association 
of Universities for Research in Astronomy, Inc., under cooperative agreement with
the National Science Foundation.} procedures. 
The TNG/HARPS-N spectra were processed with the standard user's pipeline, 
Data Reduction Software (DRS; \citealt{2002A&A...388..632P};  \citealt{2007A&A...468.1115L}).

\subsection{Radial velocities \label{Kepler}}

{The HET/HRS is a general purpose spectrograph, which is
neither temperature nor pressure-controlled. Therefore the calibration of the RV
measurements with this instrument is best accomplished with
the I$_2$  gas cell technique 
 \citep{MarcyButler1992, Butler1996}. }
Our application of this technique to HET/HRS data is described in detail in \cite{Nowak2012} and \cite{ Nowak2013}. 

The RVs from the HARPS-N were obtained with the cross-correlation  method \citep{Queloz1995, Pepe2002}.
{The wavelength calibration was done using the simultaneous Th-Ar mode of the spectrograph. The RVs were calculated by cross-correlating the stellar spectra with the digital mask for a K2 type star.}

The RV data acquired with both instruments are shown in Table \ref{RV-DATA}. { There are different zero point offsets between the data sets for every target listed in Table \ref{TableKeplerian}.}

\longtab{
\centering
\input{table_RV_BIS.tex}
}

\section{Keplerian analysis \label{results-g}}

To find the orbital parameters, we combined a global genetic algorithm (GA; \citealt{Charbonneau1995}) 
with the MPFit algorithm \citep{Markwardt2009}.  This hybrid approach is described in \cite{Gozdziewski2003, Gozdziewski2006, Gozdziewski2007}.
The  range of the Keplerian orbital parameters found with the GA was searched with the
RVLIN code \citep{WrightHoward2009}, which we modified to introduce the stellar jitter as a free parameter to be fitted in order to find the optimal solution \citep{FordGregory2007, Johnson2011}.
{ The uncertainties were estimated with the bootstrap method described by \cite{Marcy2005}.}

For a more detailed description of the Keplerian analysis presented here, we refer the reader to the first TAPAS paper
 \cite{TAPAS1}. The results of the analysis of our RV data are listed in Table \ref{TableKeplerian}.

\begin{table*}
\centering
  \caption{Keplerian orbital parameters of companions to HD 4760, BD+02 3313, TYC 0434-04538-1, and HD 96992.\label{TableKeplerian}}
\input{table_master-new}
\tablefoot{$V_0$ denotes absolute velocity of the barycenter of the system,
offset is a shift in radial velocity measurements between different telescopes,
\sjit~is stellar intrinsic jitter as defined in \cite{Johnson2011},
RMS~is the root mean square of the residuals. { T$_{0}$ is given in MJD = JD - 2400000.5.}}
\label{TableKeplerian}
\end{table*}

\subsection{HD 4760} 

HD 4760 (BD+05 109, TYC-0017-01084-1) is 
one of the least metal abundant giants in our sample, with ([Fe/H]=-0.91$\pm$0.09).

We have measured the RVs for this star at 35 epochs over about a nine year period.
Twenty-five epochs of  the HET/HRS data were obtained  between 
{ Jan 12, 2006 and Jan 22, 2013}
(2567 days, which is more than seven years). 
These data exhibit a RV amplitude of $\pm839\ms$.
We have also made additional ten observations of this star with the HARPS-N between 
{Nov 30, 2012 and June 23, 2015}
(935 days). For these observations, the measured RV amplitude is $\pm719 \ms$.   
These RV variations are more than two orders of magnitude larger than the estimated RV precision of our measurements.

\begin{figure}
   \centering
   \includegraphics[width=0.5\textwidth]{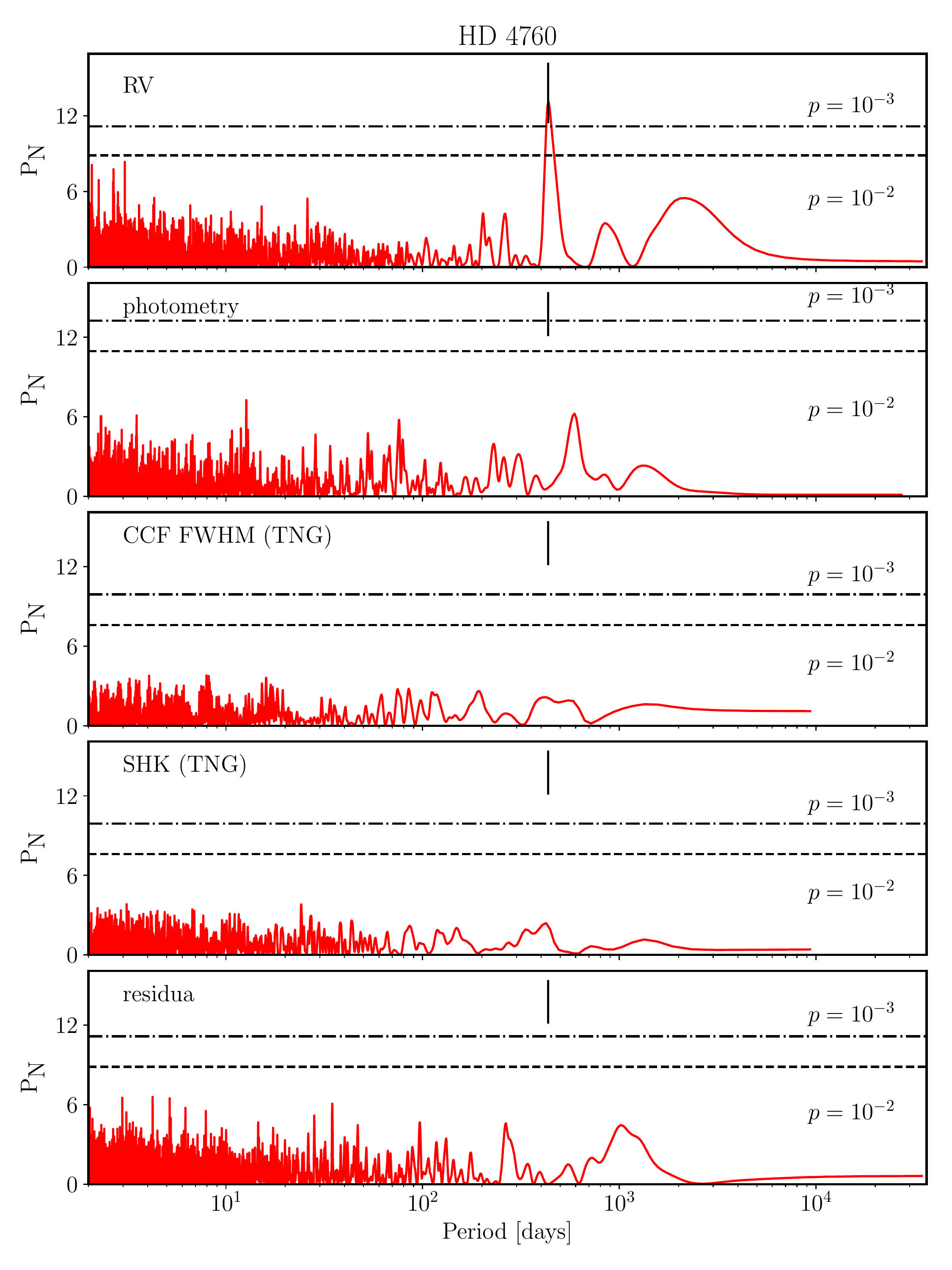}
   \caption{The Lomb-Scargle  periodogram of (top to bottom) the combined HET/HRS  and TNG/HARPS-N  RV data, the selected photometric data set, the FWHM of the cross-correlation function from TNG, the $\shk$ measured in TNG spectra and the post keplerian fit RV residua 
    for \starA.  A periodic signal is clearly present in the RV data.}
   \label{LSP_0015}
\end{figure}

The measured RVs show a statistically significant periodic signal 
{in Lomb-Scargle periodogram \citep{1976ApJSS..39..447L, 1982ApJ...263..835S, 1992nrfa.book.....P} }
(with a false alarm probability $p< 10^{-3}$) with a peak at about 430 days (Figure \ref{LSP_0015}, top panel).

\begin{figure}
   \centering
   \includegraphics[width=0.5\textwidth]{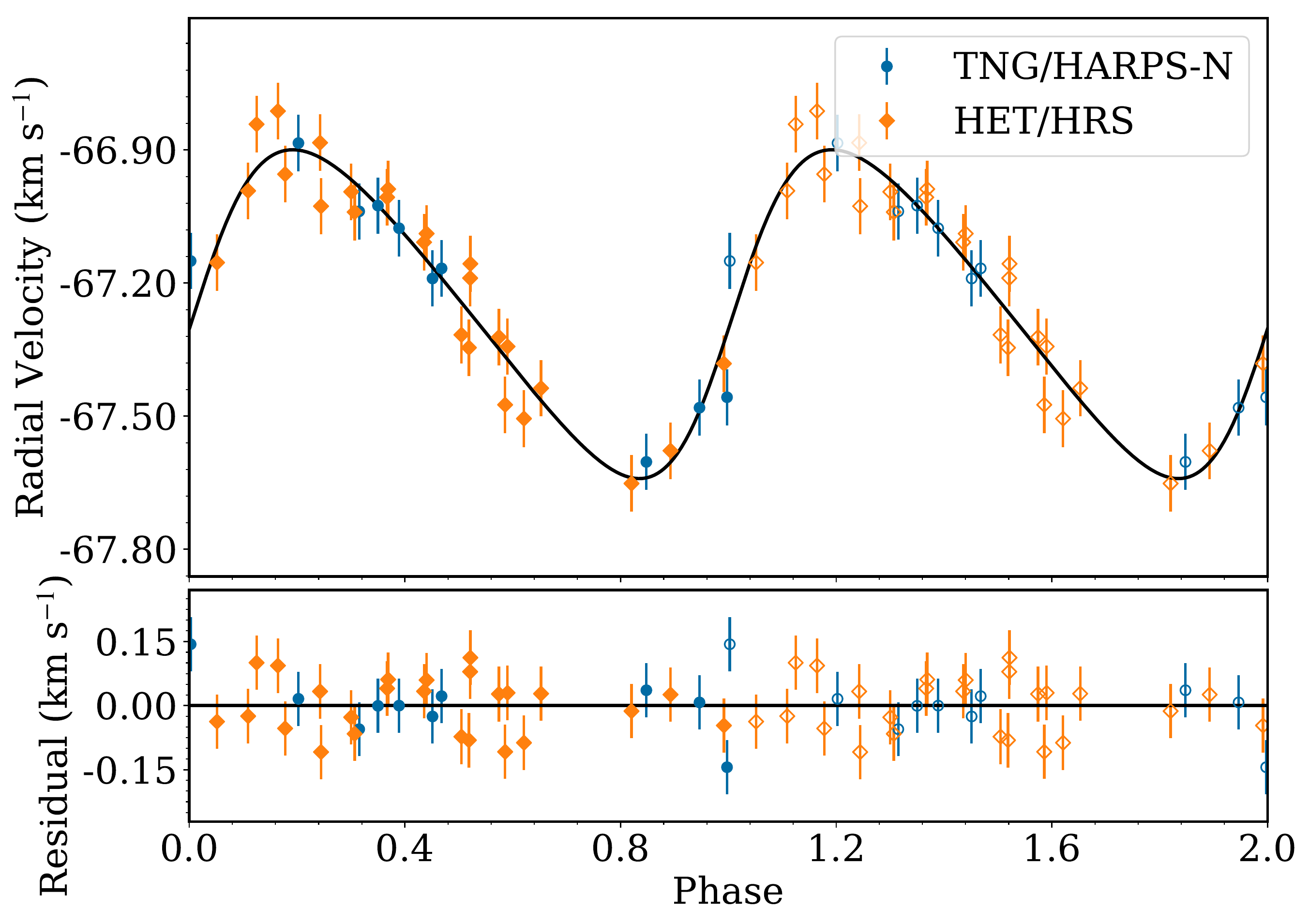}
   \caption{Keplerian best fit to combined HET/HRS (orange) and TNG/HARPS-N (blue) data for
      \starA The jitter is added to uncertainties. Open symbols denote a repetition of the data points for the initial orbital phase.}
   \label{Fit_0015}
\end{figure}

These data, interpreted in terms of a Keplerian motion, show that this star hosts a low-mass companion
on an $a=1.14$~au,  eccentric ($e=0.23$)  orbit (Figure \ref{Fit_0015}). The calculated 
minimum mass of $13.9 \pm 2.4\Mjup$, 
makes the system 
similar to the one hosted by BD+20 2457. See Table \ref{TableKeplerian} and Figure \ref{Fit_0015} for the details of the Keplerian model.

After fitting this model out of the data, the remaining RV residuals leave no trace of a periodic signal (Figure \ref{LSP_0015}, bottom panel).

\subsection{HD 96992 } 

HD 96992 (BD+44 2063, TYC 3012-00145-1) is another low-metallicity ([Fe/H]=$-0.45\pm0.08$)  giant in our sample. 

For this star, we have measured the RVs at 74 epochs over a 14 year period.
The HET/HRS data have been obtained at 52 epochs between 
{ Jan 20, 2004  and Feb 06, 2013}
(3305 days, or  $\sim$nine years), showing a maximum amplitude of $\pm157\ms$  
Twenty-four more epochs of the HARPS-N data were collected between 
{ Dec 16, 2012 and Mar 14, 2018}
(1914 days, over 5 years), resulting in a maximum RV amplitude of $\pm117 \ms$.

\begin{figure}
   \centering
   \includegraphics[width=0.5\textwidth]{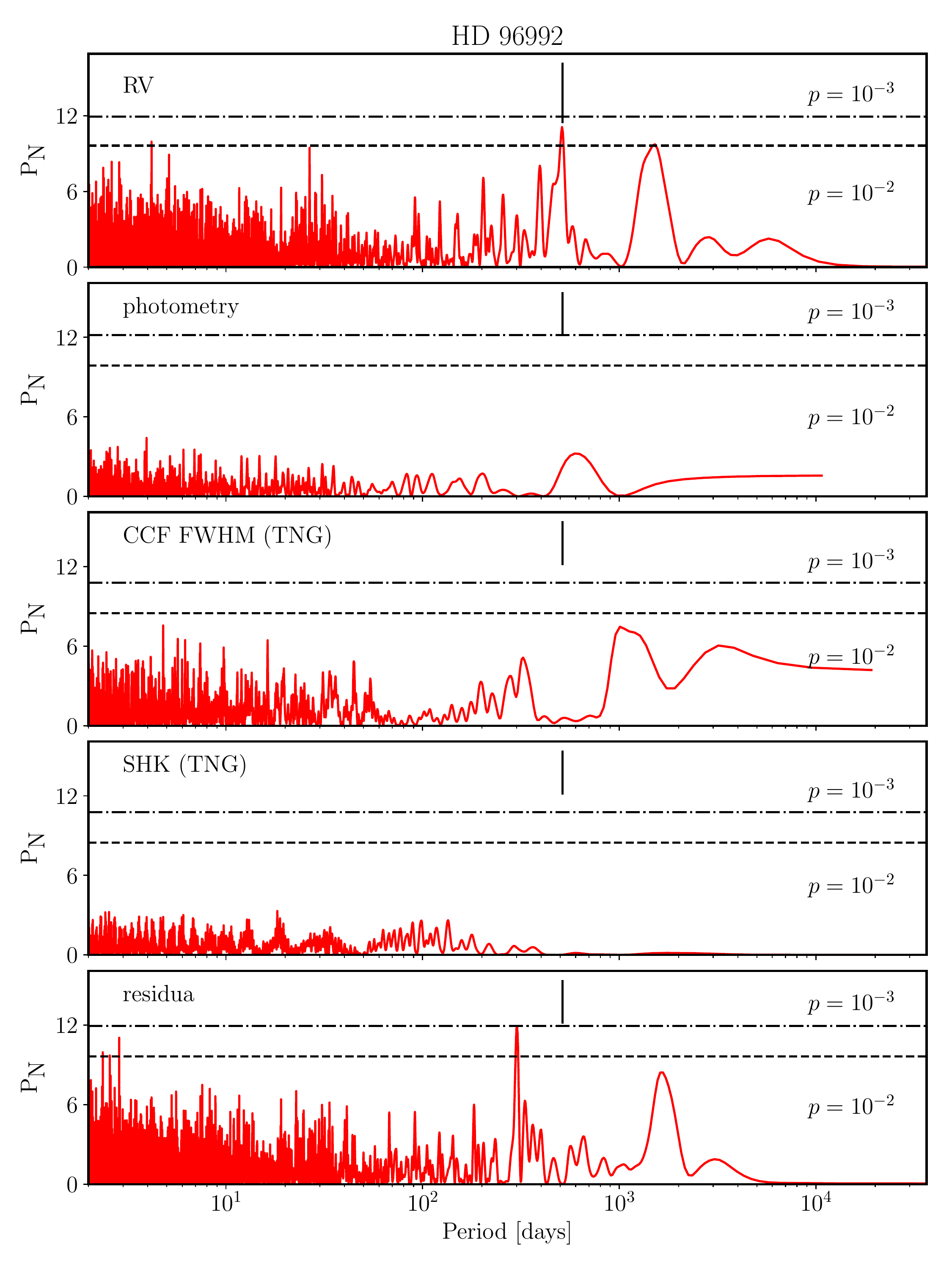}
   \caption{
   Same as Figure \ref{LSP_0015} for
      \starE. The $\approx$300 day signal in RV residuals is consistent with the estimated rotation period.}
   \label{LSP_0864}
\end{figure}

The observed maximum RV amplitude is 25-100 times larger than the estimated RV precision. 
Our RV measurements show a statistically significant periodic signal (with a false alarm probability of about $10^{-3}$) with a peak at about 510 days (Figure \ref{LSP_0864}, top panel).

\begin{figure}
   \centering
   \includegraphics[width=0.5\textwidth]{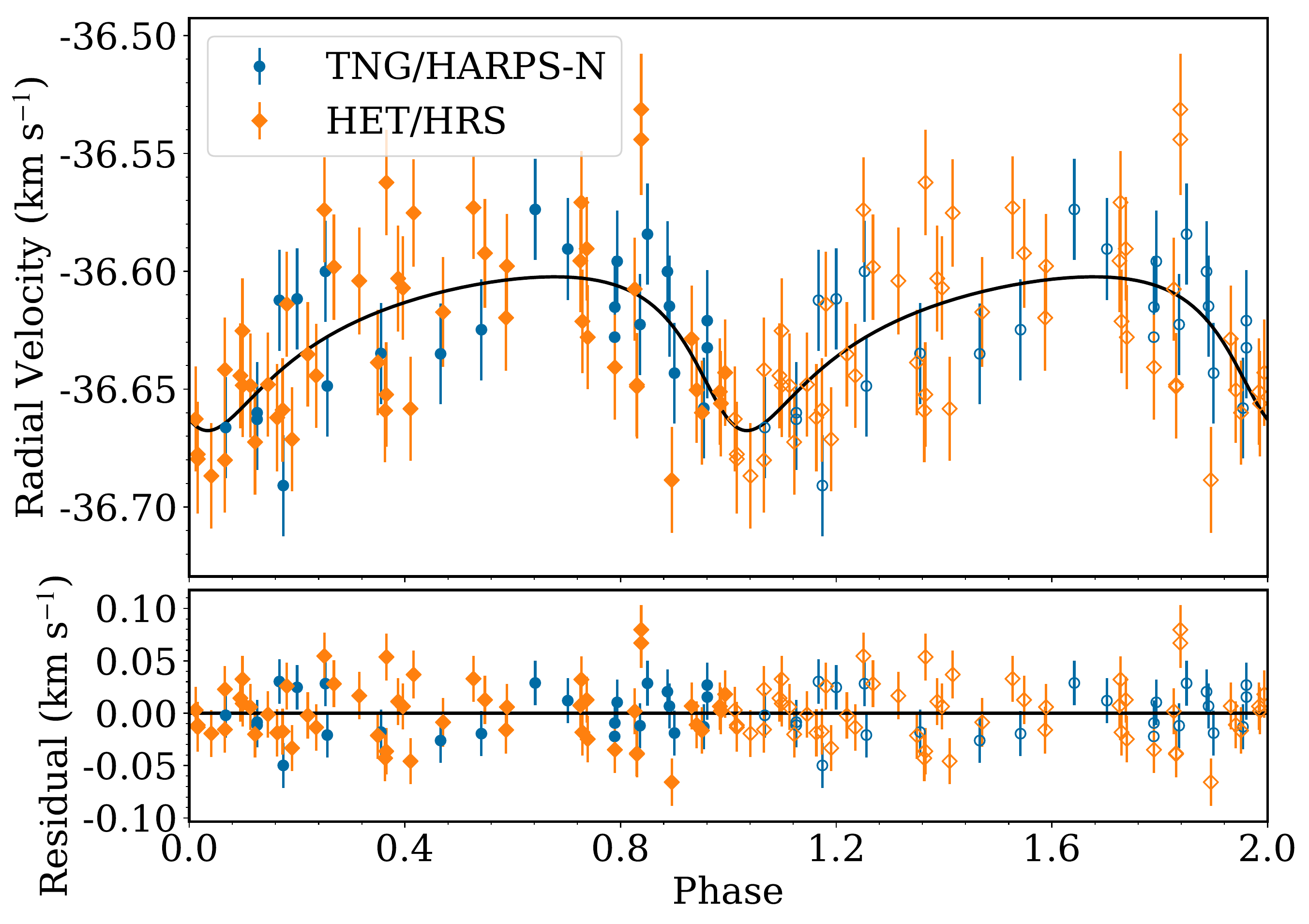}
   \caption{
   Same as Figure \ref{Fit_0015} for
      \starE. }
   \label{Fit_0864}
\end{figure}

As the result of our Keplerian model fitting to data, this single periodic RV signal suggests that HD~96992
 hosts a $m\sin i = 1.14\pm1.1 \Mjup$ mass planet on a $a=1.24$ au, rather eccentric orbit ($e=0.41$).
 The parameters of this fit are listed in Table \ref{TableKeplerian}, and Fig. \ref{Fit_0864} shows the fit to RV data.
 
As seen in Fig. \ref{LSP_0864} (bottom panel), the RV residuals, after removing the Keplerian model, reveal yet another long period signal of similar statistical significance to the 514 days one, at a period of about 300 days.

We find this periodicity consistent with our estimate of the rotation period for HD 96992. 
To test alternative scenarios for this system, we tried to model a planetary system with two planets, but the dynamical modeling with Systemic 2.16  \citep{2009PASP..121.1016M} shows that such a system is highly unstable and disintegrates after about 1000 years. 
We also attempted  to interpret the signal at 300 days as a single, Keplerian orbital motion, but it resulted in a highly eccentric orbit, and the quality of the fit was unsatisfactory. We therefore rejected these alternative solutions. 

In conclusion, we postulate that the signal at 514 days, evident in the RV data for HD 96992 is due to a Keplerian motion, and the $\sim300$ days signal remaining in the post-fit RV residuals reflects rotation of a  feature on the stellar surface.

\subsection{BD+02 3313 } 

BD+02~3313 (TYC 0405-01114-1) has a solar like metallicity of [Fe/H]=0.10$\pm$0.07,  but  it has 
27 times higher  luminosity.

\begin{figure}
   \centering
   \includegraphics[width=0.5\textwidth]{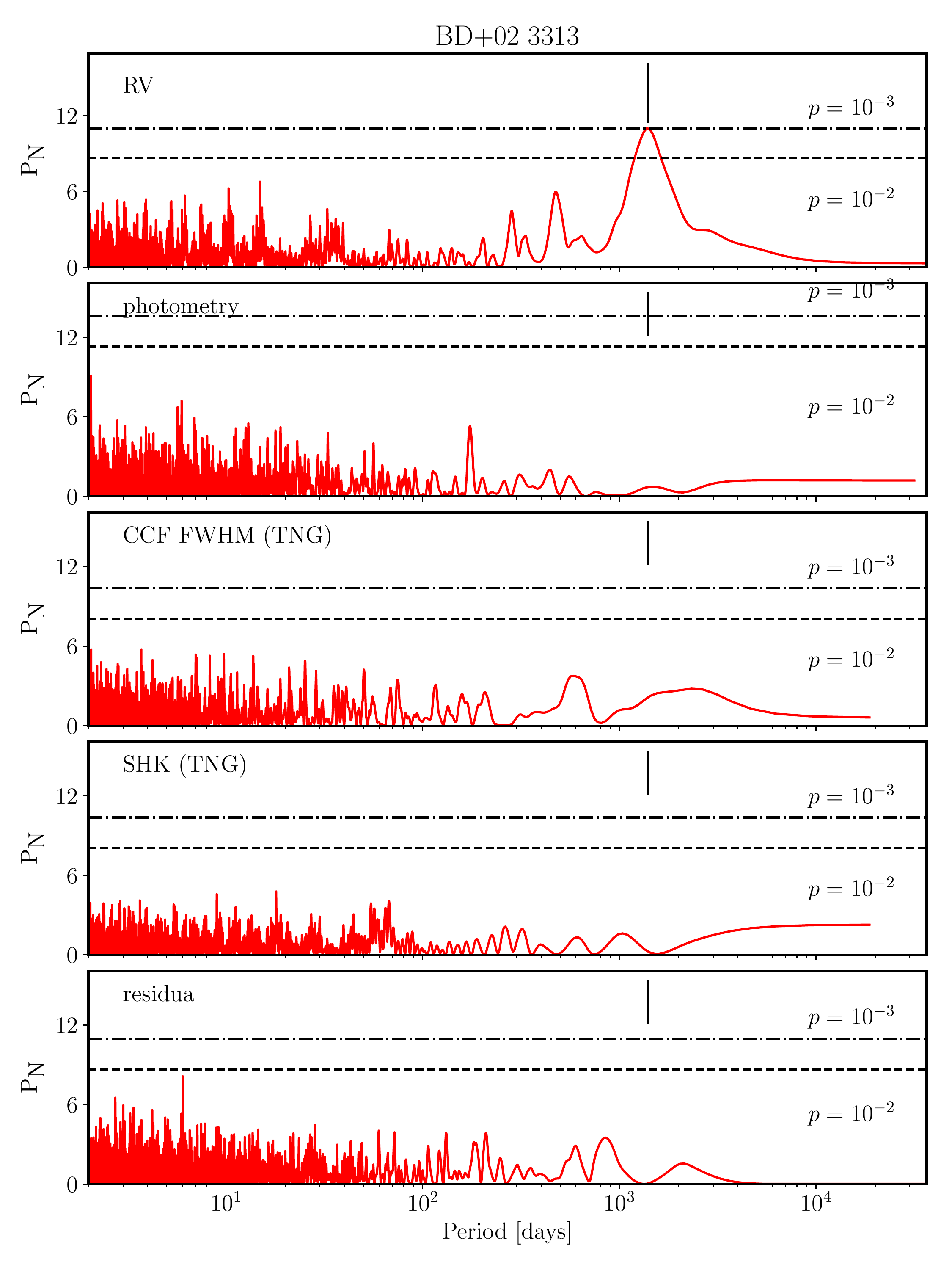}
   \caption{
   Same as Figure \ref{LSP_0015}  for
      \starB.}
   \label{LSP_0128}
\end{figure}

We have measured the RV's for this star at 29 epochs over 4264 days (11.6 years).
Thirteen epochs worth of the HET/HRS RV data were gathered  
between 
{ Jul 11, 2006 and Jun 15, 2013}
(2531 days, nearly a seven year time span). 
These RVs show a maximum amplitude of $1381\ms$. 
Additional sixteen RV measurements were made with the HARPS-N between 
{ Jan 29, 2013 and Mar 14, 2018}
(1870 day or over a 5 year time span). 
In this case, the maximum RV amplitude is $\pm1141\ms$, which is three orders of magnitude larger than the estimated RV precision.
The data show a statistically significant periodic signal (with a false alarm probability of about 10$^{-3}$) with a peak at about 1400 days (Figure \ref{LSP_0128}, top panel).

\begin{figure}
   \centering
   \includegraphics[width=0.5\textwidth]{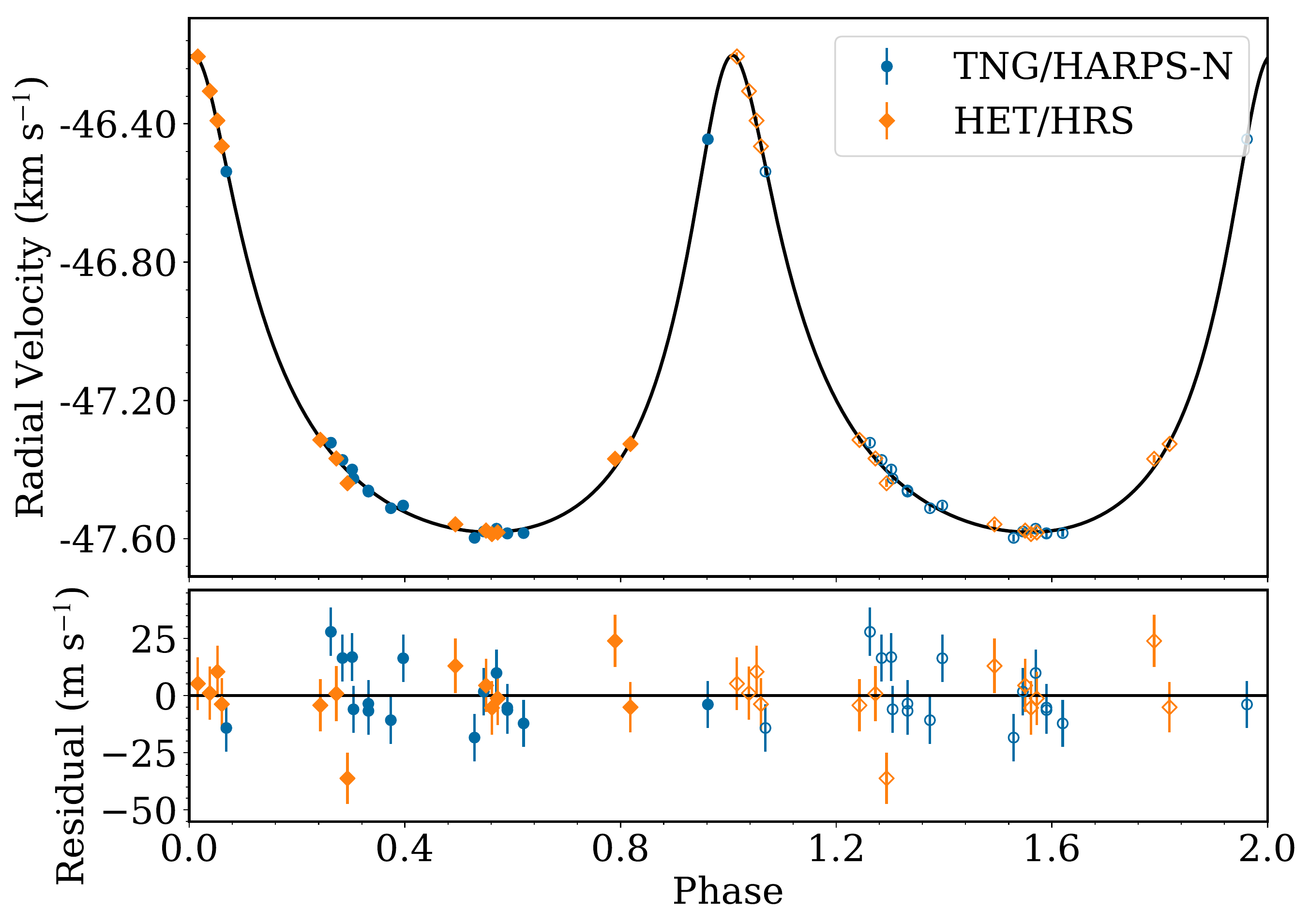}
   \caption{
     Same as Figure \ref{Fit_0015} for
      \starB.}
   \label{Fit_0128}
\end{figure}

Interpreted in terms of  the Keplerian motion, the available RVs show that this star hosts a low-mass companion, a  brown dwarf, with a minimum mass 
of $m\sin i = 34.1\pm1.1 \Mjup$. { The companion is located on a relatively eccentric orbit  ($e=0.47$),  at $a=2.47$~au, 
within the brown dwarf desert \citep{2000PASP..112..137M}, an orbital separation area below 3-5 au, known for paucity of brown dwarf companions to solar-type stars.}
Parameters of the Keplerian fit to these RV data are listed in Table \ref{TableKeplerian}, and shown in Fig. \ref{Fit_0128}.

After removing the Keplerian model from the RV data, the residuals leave no sign of any leftover periodic signal (Figure \ref{LSP_0128}, bottom panel).

\subsection{TYC 0434-04538-1 } 

TYC 0434-04538-1 (GSC 00434-04538), another  low metallicity  ([Fe/H]=-0.38$\pm$0.06) giant,   
has been observed 29 times over a period of  3557 days (9.7 years).

\begin{figure}
   \centering
   \includegraphics[width=0.5\textwidth]{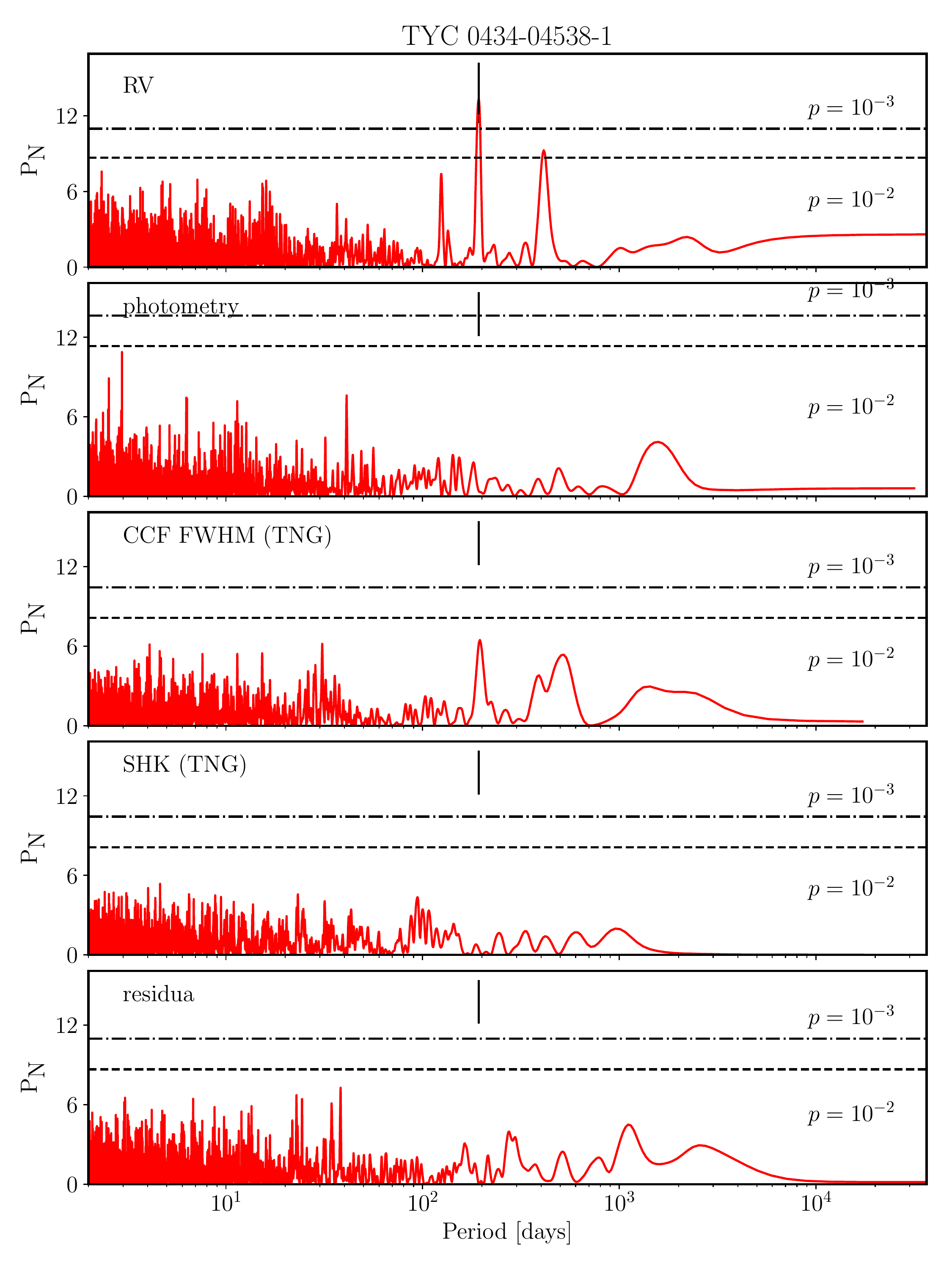}
   \caption{
   Same as Figure \ref{LSP_0015} for
      \starC. }
   \label{LSP_0154}
\end{figure}

The HET/HRS  measurements were made at twelve epochs, between 
{ Jun 23, 2008 and Jun 13, 2013}
(over 1816 days, or nearly five years), showing a maximum RV amplitude of $\pm483 \ms$ 
Additional RV measurements for this star were made with the HARPS-N at 17 eopchs between 
{Jun 27, 2013 and Mar 14, 2018}
(1721 days, 4.7 years). 
These data show a maximum RV amplitude of $\pm442 \ms$, which is similar to that seen in the HET/HRS measurements. This is over  two orders of magnitude more than the estimated RV precision.
The data show a strong, statistically significant, periodic signal
(false alarm probability $p< 10^{-3}$) with a peak at about 193 days (Figure \ref{LSP_0154}, top panel).

\begin{figure}
   \centering
   \includegraphics[width=0.5\textwidth]{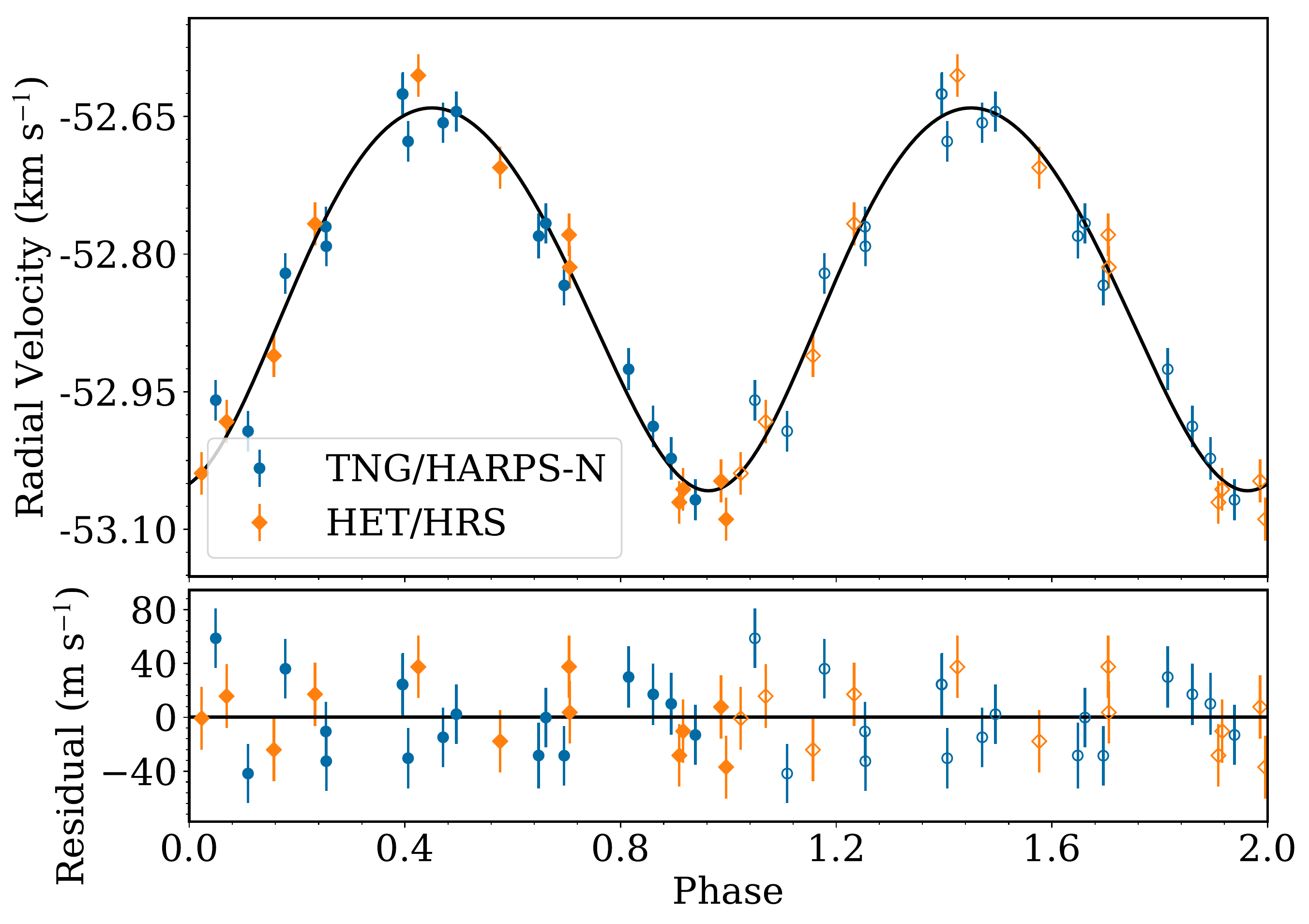}
   \caption{
    Same as Figure \ref{Fit_0015} for
      \starC. }
   \label{Fit_0154}
\end{figure}

Our Keplerian analysis shows that this star hosts a $6.1 \pm1.1 \Mjup$ 
mass planet on $a=0.66$~au, almost circular ($e=0.08$)  orbit, at the edge of the avoidance zone.  
The model parameters of the best Keplerian fit to data are presented in  Table \ref{TableKeplerian} and in Fig. \ref{Fit_0154}.

After removing this model from the observed RV measurements we do not see any other periodic signal in the periodogram of the post-fit residuals. (Figure \ref{LSP_0154}, bottom panel).

\section{Stelar variability and activity analysis\label{activity}}

{ Stars beyond the MS, especially the red giants, 
exhibit activity and various types of variability that either alter their line profiles and mimic RV shifts or cause the line shifts. 
Such phenomena, if periodic, may be erroneously taken as evidence for the presence of orbiting, low-mass companions.
}

A significant variability of the red giants has been already noted by \cite{1954AnHar.113..189P} and \cite{1989ApJ...343L..21W}, 
and made the nature of these variations a topic of numerous studies. 

All the red giants of spectral type of  K5 or later have been found to be variable photometrically with amplitudes increasing for the cooler stars \citep{1996ApJ...464L.157E, 2000ApJS..130..201H}.

Short-period RV variations in giants were demonstrated to originate from the $p$-mode oscillations by \cite{1994ApJ...432..763H}.   The first detection of multimodal oscillations in a K0 giant, $\alpha$ UMa, was published by \cite{2000ApJ...532L.133B}.

The solar-type, $p$-mode (radial) oscillations \citep{KjeldsenBedding1995, 2013MNRAS.430.2313C} are easily observable in the high precision, photometric time-series measurements,
and they have been intensely studied based on the COROT  \citep{2006ESASP1306...33B} and KEPLER \citep{2010PASP..122..131G} data, 
leading to precise mass determinations of many stars \citep{2009Natur.459..398D, 2010ApJ...713L.176B, 2010A&A...522A...1K, 2011MNRAS.414.2594H}.
\cite{2020MNRAS.493.1388Y} present an HRD with the amplitudes and frequencies of solar-like oscillations from the MS up to the tip of the RGB.

The granulation induced ,,flicker''  \citep{2017A&A...605A...3C, 2019ApJ...883..195T},
with characteristic time scales of $\approx$hours, is undoubtedly an additional unresolved component to the RV scatter in the red giants.

Low amplitude, non-radial oscillations (mixed modes of \citealt{2001MNRAS.328..601D})  in the red giants 
(with frequencies of $\approx$5-60 cycles per day) were first detected by \cite{2006A&A...458..931H}.
 They were later unambiguously confirmed  using the COROT data by
\cite{2009Natur.459..398D}, who also  found that the lifetimes of these modes are on the order of a month.

With the typical timescales for the red giants, ranging from hours to days, the short-period variations typically remain unresolved in low-cadence 
observations, focused on the long-term RV variations, and they contribute an additional uncertainty to the RV measurements.

In the context of planet searches, long period variations of the red giant stars  
are more intriguing, because they may masquerade as the low-mass companions.
Therefore, to distinguish between 
line profile shifts due to orbital motion 
from those caused by, for instance, pulsations, 
and line profile variations induced by stellar activity,
it is crucial to understand processes that may cause the observed line shifts 
by studying the available  stellar activity indicators.

As the HET/HRS spectra do not cover the spectral range, where Ca II H \& K  lines are present, we  use the line bisector, and the H$_{\alpha}$ line index as activity indicators.
In the case of TNG HARPS-N spectra, in addition to the RVs, the DRS provides  the FWHM  of the cross-correlation function (CCF) between 
 the stellar spectra and the digital mask and the line bisector (as defined in \citealt{2001A&A...379..279Q}), both being sensitive activity indicators.

\subsection{Line bisectors}

The spectral line bisector (BIS) is a measure of the asymmetry of a spectral line, which can arise for such reasons as blending of lines, a 
surface feature (dark spots, for instance), oscillations, pulsations, and granulation (see \citealt{2005PASP..117..711G} for a discussion of BIS properties).
BIS has been proven to be a powerful tool to detect 
starspots  and background binaries \citep{2001A&A...379..279Q, 2002A&A...392..215S}
that can mimic a planet signal in the RV data.  { In the case of surface phenomena (cool spots), the anti-correlation between BIS and RV is expected \cite{2001A&A...379..279Q}. }
In the case of a multiple star system with a separation smaller than that of the fiber of the spectrograph, the situation is more complicated: a correlation, anti-correlation, or lack of correlation may occur, depending on the properties of the components (see \citealt{2015MNRAS.451.2337S} and \citealt{2018MNRAS.478.4720G} for a discussion).
Unfortunately,   for the slow-rotating giant stars, like our targets,  BIS is not a sensitive activity indicator  \citep{2003A&A...406..373S, 2014A&A...566A..35S}.

{ The HET/HRS and the HARPS-N bisectors are defined differently and were calculated from different instruments and spectral line lists. They are not directly comparable
and have to be considered  separately. 
All the HET/HRS spectral line bisector  measurements  were obtained from the spectra used for  the I$_2$ gas-cell technique  \citep{MarcyButler1992, Butler1996}. 
The combined
stellar and iodine spectra were first cleaned of the I$_2$  lines by
dividing them by the corresponding iodine spectra imprinted
in a flat-field ones, and then cross-correlated with a binary 
K2 star mask.  A detailed description of this procedure is described in \cite{ Nowak2013}.
{ As stated in Sect. \ref{Kepler} , HET/HRS is not a stabilized spectrograph, and the lack of correlation for BIS should be treated with caution, 
as it might be a result of the noise introduced by the varying instrumental profile. }

The Bisector Inverse Slopes  of the  cross-correlation functions
from the HARPS-N data were obtained with the \cite{2001A&A...379..279Q} method,
using the standard HARPS-N user's pipeline, 
which utlilizes  the simultaneous Th-Ar calibration mode of the spectrograph  and  the cross-correlation  
mask with a stellar spectrum (K2 in our case).

In all the cases presented here, the RVs do not correlate with the line bisectors at the accepted significance level (p=0.01),
 see Tables \ref{HET_actv} and \ref{TNG_actv}.
We conclude, therefore, that the HET/HRS and the HARPS-N BIS RV data have not been
influenced by spots or background binaries.}

\subsection{The $\ha$ activity index}

The  H$_{\alpha}$ line is a widely used indicator of  the chromospheric activity, as the core of this line
is formed in the chromosphere. The increased stellar activity shows a correspondigly filled H$_{\alpha}$ profile.
Variations in the flux in the line core can be measured with  the 
 $\ha$ activity index,  defined as  the flux ratio in a band centered on the H$_{\alpha}$ to the flux in the reference bands. 
 We  have measured the H$_{\alpha}$ activity index ($\ha$) in both the HET/HRS and the TNG/HARPS-N spectra
using the procedure described  in \cite{2013AJ....146..147M}
(cf. also \citealt{2012A&A...541A...9G} or \citealt[][and references therein]{2013ApJ...764....3R}). 

The HET/HRS spectra were obtained with the use of the iodine cell technique  meaning that the iodine spectrum was imprinted on the stellar one. 
To remove the weak iodine lines in the H$_{\alpha}$ region, we divided an order of spectrum by the H$_{\alpha}$
by the corresponding order of the GC flat spectrum, before performing the $\ha$ index analysis. 

A summary of our $\ha$ analysis in the HET/HRS data is shown in  Table \ref{HET_actv}, 
and a summary of the HARPS-N $\ha$ data analysis is presented in Table \ref{TNG_actv}.  
No statistically significant correlation between $\ha$ and the RV data
 has been found for our sample stars.

\subsection{Calcium H \& K doublet}

 The reversal profile in the cores of Ca H and K lines, i.e., the emission structure 
 at the core of the Ca absorption lines, is another commonly used indicator of stellar activity \citep{EberhardSchwarzschild1913}. 
The Ca II H \& K lines are located 
at the blue end of the TNG/HARPS-N spectra, which is the region with the lowest S/N for our red targets.
The S/N of the spectra for the stars discussed here varies between 2 and 10.
Stacking the spectra to obtain a better S/N is not possible here as they have been taken at least a month apart. 
For every epoch's usable spectrum for a  given star, we calculated the $\shk$ index following 
the formula of \cite{Duncan1991}, and we calibrated it against the Mount Wilson scale with the formula provided in \cite{2011arXiv1107.5325L}. 
We also searched the $\shk$ indices for variability and found none (see periodograms in Figures \ref{LSP_0015}, \ref{LSP_0864}, \ref{LSP_0128} and \ref{LSP_0154}).
Therefore, we conclude that the determined $\shk$ indices are not related to the observed RV variations.

\subsection{Photometry}

Stellar activity and pulsations can also  manifest themselves  through changes in the brightness of a star. All our targets have been observed by 
large photometric surveys. We collected the available data for them from several different catalogs: ASAS \citep{1997AcA....47..467P}, NSVS \citep{2004AJ....127.2436W},
Hipparcos \citep{1997ESASP1200.....P} and SuperWASP \citep{2006PASP..118.1407P}.  
We then selected the richest and the most precise data set from all available ones for a detailed variability and period search.
The original photometric time series were binned by one day intervals. 
We found no periodic signal in the selected time-series photometry for any of our targets
 (see periodograms in Figures \ref{LSP_0015}, \ref{LSP_0864}, \ref{LSP_0128} and \ref{LSP_0154}).
Table \ref{photo} summarizes the results for the selected data.

\subsection{CCF FWHM}
{ The stellar activity  and surface phenomena impact} the shape of the lines in the stellar spectrum. 
Properties of CCF, 
a mean profile of all spectral lines, are used as activity indicators. 
In a recent paper  \cite{2017A&A...606A.107O} found the CCF FWHM to be the best indicator of stellar activity available from the HARPS-N DRS {(for main sequence
sun-like stars)}, in accordance with the previous results of \cite{2009A&A...506..303Q} and \cite{2011MNRAS.411.1953P}. 
These authors recommend it to reconstruct the stellar RV jitter as the CCF FWHM correlates well with the activity-induced RV in the stars of various activity levels.

\begin{figure}
   \centering
   \includegraphics[width=0.5\textwidth]{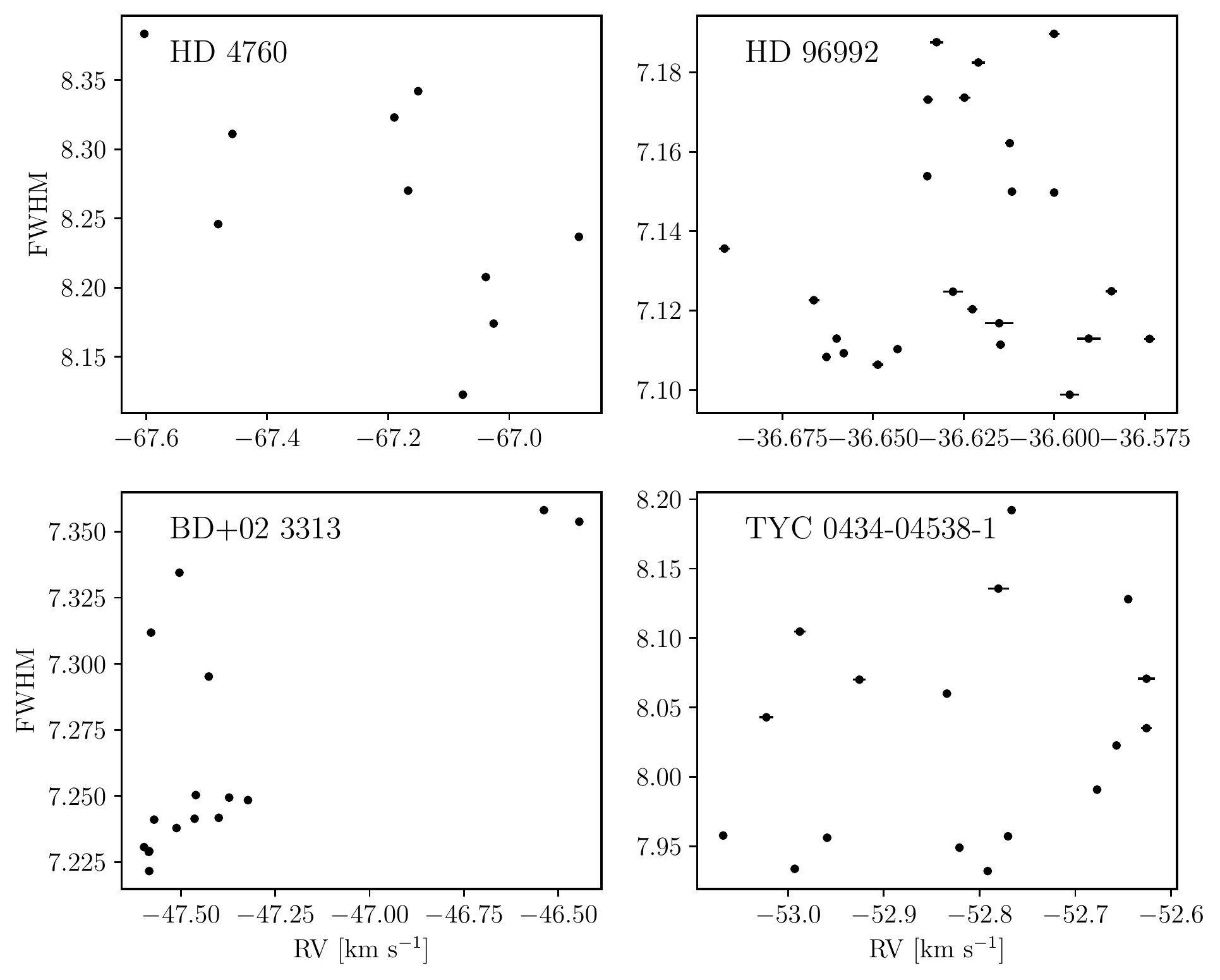}
   \caption{Radial velocities plotted against cross-correlation function FWHM  for TNG/HARPS-N data.}
   \label{rv_fwhm_fig}
\end{figure}

For all the HARPS-N observations available for our targets, we have correlated the FWHM of the CCF against 
the RV measurements for the TNG/HARPS-N data set. The presence of a correlation means that the observed variability may stem from distorted  spectral lines, possibly due to stellar activity.
The results of this analysis are shown in Table \ref{TNG_actv}  and in Fig. \ref{rv_fwhm_fig}. In the case of  BD+02 3313 
we found a statistically significant ($\mathrm{r}=0.73 >\mathrm{r}_{c}=0.62$) correlation at the accepted confidence level of p=0.01 between the observed RV and the CCF FWHM.

We also searched the CCF FWHM from HARPS-N   for variability but found no statistically significant signal (see periodograms in Figures \ref{LSP_0015}, \ref{LSP_0864}, \ref{LSP_0128} and \ref{LSP_0154}).

\begin{table*}
\centering
  \caption{Summary of the activity analysis. Observations span (OS) is the total observation span covered by HET and TNG, $K_{\mathrm{osc}}$ is an amplitude of expected 
  solar-like oscillations \citep{KjeldsenBedding1995}, OS$_{HET}$ is a observing periods for HET only, $K$ denotes an
  amplitude of observed radial velocities defined as $RV_{\mathrm{max}} -RV_{\mathrm{min}} $,
  $\overline{\sigma_{\mathrm{RV}}}$ is an average RV uncertainty.
All  linear correlation coefficients  are calculated with reference to RV. The last columns provides the number of epochs. 
  }
  \small
\begin{tabular}{l | c S[table-format=1.1] |
    c S[table-format=1.1] S[table-format=1.1] | 
    S[table-format=1.1] S[table-format=1.1] | S[table-format=1.1]S[table-format=1.1] |
    S[table-format=1.1] S[table-format=1.1] | S[table-format=1.1] S[table-format=1.1] | c}
\hline 
\multirow{3}{*}{Star} &   & &&&&\multicolumn{5}{c}{HET/HRS}  \\

   & OS & $K_{\mathrm{osc}}$ &
   OS$_{HET}$ &$K$ & $\overline{\sigma_{\mathrm{RV}}}$ & \multicolumn{2}{c|}{BIS}

   &\multicolumn{2}{c|}{$\ha$} 
  &No\\
   &[days]&[$\ms$]&[days]&[$\ms$]&[$\ms$]&r&p&r&p&\\
\hline   
          HD 4760 &   3449 & 189.68 &   2567 &   839.01 &   6.70 &   0.18 &   0.38 &  0.05 & 0.81  & 25 \\
         HD 96992 &   5167 &   7.19 &   3305 &   157.24 &   6.29 &   0.04 &   0.76 &  0.13 & 0.36  & 52 \\
       BD+02 3313 &   4264 &   6.26 &   2531 &  1381.25 &   5.32 &  -0.20 &   0.51 &  0.24 & 0.45  & 13 \\
 TYC 0434-04538-1 &   3551 &  10.52 &   1816 &   483.64 &   8.06 &   0.26 &   0.41 &  0.28 & 0.40 & 12 \\

\hline
\end{tabular}
\label{HET_actv}
\end{table*}

\begin{table*}
\centering
  \caption{Summary of the activity analysis. The  OS$_{TNG}$ is an observing period for TNG only,   $K$ denotes an
  amplitude of observed radial velocities defined as $RV_{\mathrm{max}} -RV_{\mathrm{min}} $,
  $\overline{\sigma_{\mathrm{RV}}}$ is an average RV uncertainty.
All  linear correlation coefficients  are calculated with reference to RV. 
The last columns provides the number of epochs.
  }
  \small
\begin{tabular}{l | c S[table-format=1.1]  S[table-format=1.1] | 
    S[table-format=1.1] S[table-format=1.1] | S[table-format=1.1]S[table-format=1.1] |
    S[table-format=1.1] S[table-format=1.1] | S[table-format=1.1]  S[table-format=1.1]  |  c}
\hline 
  \multirow{2}{*}{Star} & 

   OS$_{TNG}$ &$K$ & $\overline{\sigma_{\mathrm{RV}}}$ & \multicolumn{2}{c|}{BIS} & \multicolumn{2}{c|}{FWHM} 
   &\multicolumn{2}{c|}{$\ha$} 
   &\multicolumn{2}{c|}{$\shk$}  &No\\
   &[days]&[$\ms$] 
   & $[\ms]$&r&p&r&p&r&p&r&p&\\
\hline   
          HD 4760 &   
           934 &   718.30 &   1.10 &   0.56 &   0.09  &  -0.61 &   0.06  &   0.71 &   0.02  &  -0.16 &   0.66  & 10\\
         HD 96992 &   
            1914 &   117.15 &   1.65 &  -0.24 &   0.25  &   0.10 &   0.63  &   0.30 &   0.15  & 0.09 &   0.69  & 24\\
       BD+02 3313 &   
       1870 &  1153.10 &   1.54 &  -0.46 &   0.08  &   0.73 &   0.00  &   0.57 &   0.02  &   0.16 &   0.55  & 16\\
 TYC 0434-04538-1 &   
 1721 &   442.23 &   4.53 &   0.50 &   0.04  &   0.27 &   0.29  &   0.59 &   0.01  &  -0.19 &   0.47  & 17\\

\hline
\end{tabular}
\label{TNG_actv}
\end{table*}

\begin{table*}
\centering
\caption{A summary of long photometric time-series available for presented stars.}
  \begin{tabular}{llllll}
  \hline
 			& HD 4760 			& HD 96992 		& BD+02 3313 &	 TYC 0434 04538 1 \\	
\hline
\hline
Source 		& ASAS				& Hipparcos 		& ASAS			& ASAS			\\ 
to [HJD]		&2455168.56291		& 2448960.99668	& 2455113.52337	& 2455122.52181	\\ 
N points		& 288				& 96				& 414			& 419			\\ 
filter			& V					& Hp				& V				& V				\\ 
mean mag.	& 7.483				& 8.741			& 9.477			& 10.331			\\ 
rms mag.		& 0.023				& 0.019			& 0.018			& 0.019			\\ 
\hline
\end{tabular}
\label{photo}
\end{table*}

\section{Discussion. }

{ 
\cite{HatzesCochran1993} have suggested  that the low-amplitude, long-period RV variations in red giants are attributable to pulsations, 
stellar activity - a spot rotating with a star, or low-mass companions. }
Such RV variations  have been successfully demonstrated to be due to a presence of low-mass companions to many giants. Starting from $\iota$ Dra 
 \citep{2002ApJ...576..478F}, 112 giants with planets have been listed in the compilation by Sabine Reffert - https://www.lsw.uni-heidelberg.de/users/sreffert/giantplanets/giantplanets.php. For some giants, however, the companion hypothesis has been debatable.

The nature of the observed RV  
long-term variability 
in some giants \citep{1933BHarO.893...19O, 1954AnHar.113..189P, 1963AJ.....68..253H} remains a riddle.  
Long, secondary period (LSP) photometric variations of AGB stars but also  the luminous red giant (K5-M) stars
near the tip of the  first giant branch (TFGB),
brighter than logL/L$_{\odot}$$\sim$2.7, 
 were detected in MACHO  \citep{1999IAUS..191..151W}: their sequence D in the period-luminosity relation for the variable semi-regular giants), and in OGLE  \citep{2007ApJ...660.1486S, 2009AcA....59..239S, 2011AcA....61..217S, 2013AcA....63...21S} data.  
They associate primary (but not always stronger)  pulsations in these stars with typically $\approx$10 times shorter periods (usually on sequence B, first overtone pulsations,  of \citealt{1999IAUS..191..151W}).
Depending on the adopted detection limit, 30-50$\%$ of luminous red giants may display LSP \citep{2007AcA....57..201S}.
With photometric amplitudes of the order of 1 mag, periods ranging from 250 to 1400 days, and RV amplitudes of 2-7 $\kms$ \citep{2004ApJ...604..800W, 2009MNRAS.399.2063N}, LSP in luminous giants should be easily detectable in precise RV planet searches. 

\cite{2004AcA....54..129S}, following suggestions by \cite{2002MNRAS.337L..31I} and \cite{2003MNRAS.343L..79K},
demonstrated that in the LMC, LSP are also present in stars below the TFGB, in the first ascent giants. 
These stars, OGLE Small Amplitude Red Giants (OSARGs, \citealt{2004MNRAS.349.1059W}), show much lower amplitudes ($<0.13$mag in I band). 

The origin of LSP is practically unknown. 
Various scenarios: the eccentric motion of an orbiting companion of mass $\approx0.1\Msun$,  
radial and nonradial pulsations, rotation of an ellipsoidal-shaped red giant, episodic dust ejection,
and starspot cycles, were discussed in \cite{2004ApJ...604..800W}. 
These authors propose a composite effect of large-amplitude non-radial, g+ mode pulsation, and strong starspot activity as the most feasible model.
 \cite{2014ApJ...788...13S}  proposed another scenario, a 
low-mass companion in a circular orbit just above the
surface of the red giant,  followed by a dusty cloud that regularly obscures the giant and causes the apparent luminosity variations.
 More recently, \cite{2015MNRAS.452.3863S} proposed oscillatory convective modes as another explanation for the LSP. 
 Thse models, however, cannot explain effective temperatures of AGB stars ($\log L/\Lsun\ge 3$, $M/\Msun=2$) 
and periods at the same time.

Generally, the observational data seem to favour binary-type scenarios for  LSP in giants, as for shorter periods the sequence D coincides with the E sequence of \cite{1999IAUS..191..151W},  
formed by close binary systems, in which one of the components is a red giant deformed due to the tidal force \citep{2004AcA....54..347S, 2012MNRAS.421.2616N}.
Sequence E  appears then, to be an extension of the D sequence towards  lower luminosity giants \citep{2004AcA....54..347S}, and some of the LSP cases may be explained by ellipsoidal variability (ibid.).  See \cite{2012MNRAS.421.2616N} for a discussion of differences of properties of pulsating giants in  sequences D and E.

Recently, 
\cite{2014A&A...566A.124L},
\cite{2016AJ....151..106L}, and 
 \cite{ 2018A&A...619A...2D}, 
 invoked LSP as a potential  explanation of observed RV variations in 
HD 216946 (M0 Iab var, logg=0.5$\pm$ 0.3, R=350R$_{\odot}$, M=6.8$\pm$1.0 M$_{\odot}$), 
$\mu$ UMa (M0 III SB, T$_{eff}$=3899 $\pm$35K, logg=1.0, M=2.2M$_{\odot}$, R=74.7R$_{\odot}$, L=1148L$_{\odot}$);
and  NGC 4349 No. 127 (L=504.36L$_{\odot}$, logg=1.99$\pm$0.19, R=36.98$\pm$4.89R$_{\odot}$, M=3.81$\pm$0.23M$_{\odot}$),
respectively. 


An interesting case of Eltanin ($\gamma$ Dra), a giant  with RV variations that disappeared after several years, was recently discussed by \cite{2018AJ....155..120H}. 
This  $M = 2.14 \pm 0.16 \Msun$ star, ($ R = 49.07 \pm 3.75 \Rsun$,  and  $L=510 \pm 51 \Lsun$ , op. cit. and  [Fe/H] = +0.11 $\pm$  0.09, $\Teff = 3990 \pm 42$ K, and $\log g =1.669 \pm 0.1$ \citealt{2012A&A...538A.143K}) 
exhibited  periodic RV variations that mimicked an $m \sin i = 10.7 \Mjup$ companion in 702 day orbit between 2003 and 2011. 
In the more recent data, collected between 2011 and 2017, these variations disappeared. The nature of this type of variability is unclear. 
The authors suggest a new form of stellar variability, possibly related to oscillatory convective modes \citep{2015MNRAS.452.3863S}.

Aldebaran ($\alpha$ Tau) was studied in a series of papers
\citep{1993ApJ...413..339H, 1998MNRAS.293..469H}
in search for the origin of observed long-term RV variations. \cite{2015A&A...580A..31H}, based on 30 year long observations,
put forward a planetary hypothesis to this $M= 1.13 \pm 0.11 \Msun$ giant star
($\Teff = 4055 \pm 70$ K, $\log g=1.20 \pm 0.1$, and [Fe/H]=-0.27 $\pm$ 0.05,
$R=45.1 \pm 0.1 \Rsun$,  op.cit.).
They proposed a $m \sin i=6.47 \pm 0.53 \Mjup$ planet in 629 day orbit and a 520 day rotation modulation by a stellar surface structure.
Recently, \cite{2019A&A...625A..22R} showed, that 
in 2006/2007, the statistical power of the $\approx620$ day period exhibited a temporary but significant decrease.
They also noted an apparent phase shift between the RV variations and orbital solution at some epochs.
These authors note the resemblance of this star and $\gamma$ Dra, and also point
to oscillatory convective modes  of \cite{2015MNRAS.452.3863S} as the source of observed variations.

Due to the unknown underlying physics of the LSP, these claims are difficult to dispute.
However, a mysterious origin of the LSP certainly makes luminous giants very intriguing  objects, especially in the context of searching for low-mass companions around them.

Another phenomenon that can mimic low-mass companions in precise RV measurements is 
starspots rotating with the stellar disk. They can affect spectral
line profiles of magnetically active  stars and mimic periodic RV variations caused by orbiting companions \citep{Vogt1987, Walker1992, SaarDonahue1997}. 

Slowly rotating, large G and K giants, are not expected to exhibit strong surface magnetic fields. 
Nevertheless, they may show activity in the form of emission in the cores of strong chromospheric lines,
photometric variability, or X-ray emission  \citep{2014IAUS..302..350K}.
 In their study of 17 377 oscillating red giants from Kepler \cite{2017A&A...605A.111C} identified only 2.08$\%$ of the stars
to show a pseudo-periodic photometric variability likely originating from surface spots (a frequency consistent with the fraction of spectroscopically detected, rapidly rotating giants in the field).

The most extreme example of a slowly rotating giant with a relatively strong magnetic field of 100~G  \citep{2008A&A...491..499A} is EK Eri.
This star was found to be a $14 \Lsun$,  $1.85\Msun$  GIV-III giant with $\Teff=5125$ K, $\log g=3.25$ and photometric period of $306.9\pm0.4$ days by \cite{1999A&A...343..175S}.
A detailed spectroscopic study by \cite{2005A&A...444..573D} has shown RV variations of about $100\ms$ with the rotation period and a positive correlation between RV and BIS. 
In a following extensive study of this object, \cite{2010A&A...514A..25D} constrain the atmospheric parameters, suggest that the rotation period is twice the photometric period $P_{\mathrm{rot}} = 2P_{\mathrm{phot}}=617.6$ days, and present a 1979-2009 V photometry time series. The amplitude of the periodic variations is about 0.3 mag.

Another example is Pollux ($\beta$ Gem), a  slowly rotating M=2.3$\pm$0.2M$_{\odot}$ \citep{2019Ap.....62..338L} giant with a detected magnetic field.
In a series of papers: \cite{HatzesCochran1993, 1993PASP..105..825L, 2006ApJ...652..661R}, it has been found to have a planetary-mass companion in $589.7\pm3.5$ days orbit, and no sign 
of activity. 
Later on, that result was confirmed  by \cite{2006A&A...457..335H}, who also estimated the star's rotation period to be 135 days.
\cite{2009A&A...504..231A} detected a weak magnetic field of -0.46$\pm$0.04 G in Pollux, and \cite{2014IAUS..302..359A} 
proposed a two-spot model that might explain the observed RV variations. However, in their model 
,,photometric variations of larger amplitude than those detected in the Hipparcos data were predicted''.
{ In their recent paper \cite{2021arXiv210102016A} find that the 
longitudinal magnetic field of Pollux  varies with a sinusoidal behaviour and a period of 660$\pm$15 days, similar, to that of the RV variations  but different. }

The presence of spots on a stellar surface may mimic low-mass companions, if the spots show a similar,
repetitive pattern for a sufficiently long period of time. However, very little is known about the lifetime of spots on the surface of single inactive, slowly rotating giants.
\cite{2009A&A...506..245M} estimate that on the surface of F-G type, MS stars spots may form for a duration of 0.5-2 times the rotation period.
\cite{1990AJ....100.2017H} studied  the evolution of four spots on the surface of a long-period RS CVN binary V1817 Cygni (K2III), 
and estimated their lifetimes to be two years.
Also, \cite{2006PASP..118.1112G} identified  a magnetically active region on the surface of Arcturus (K2 III) that lasted for a period of 2.0$\pm$ 0.2 yr (the star was found to present a weak magnetic field by \citealt{2011A&A...529A.100S}).
\cite{2008ApJ...679.1531B} have published observations that suggest  migration of an active region on the surface of Arcturus over a timescale of 115-253 days.
A similar result, suggesting a 0.5-1.3 year recurrence period in starspot emergence,  was derived in the case of a rapidly rotating K1IV star, KIC 11560447 \citep{2018MNRAS.474.5534O}.

The lifetime of spots on surfaces of single, low activity giants appears to be on the order of $\sim$2 years. 
 A long enough series of data, covering several lifetimes of starspots is clearly required to rule out or confirm activity-related features as the origin of the observed RV variability.

\subsection{HD 4760}

HD 4760 is the most luminous and one of the most evolved stars in our sample with $\llsun=2.93\pm0.11$. 
Its large radius ($R/\Rsun=42\pm9$), low metallicity ([Fe/H]=-0.91$\pm$0.09), and small $\log g =1.62\pm0.08$ 
make it a twin to BD+20 2457, taking into account the estimated uncertainties. 

The observed RV amplitude is  about four times larger than the expected amplitude of the $p$-mode oscillations (cf. Table \ref{HET_actv}). 
We find the actual RV jitter ($\sigma_{jitter}$) in HD 4760 about three times smaller than the expected ($K_{\mathrm{osc}}$) from the p-mode oscillations (Table \ref{TableKeplerian}). Such discrepancy  cannot be explained by the estimated uncertainties, and it suggests that  they may have been underestimated in either luminosity or mass (or both).

\begin{figure}
   \centering
   \includegraphics[width=0.5\textwidth]{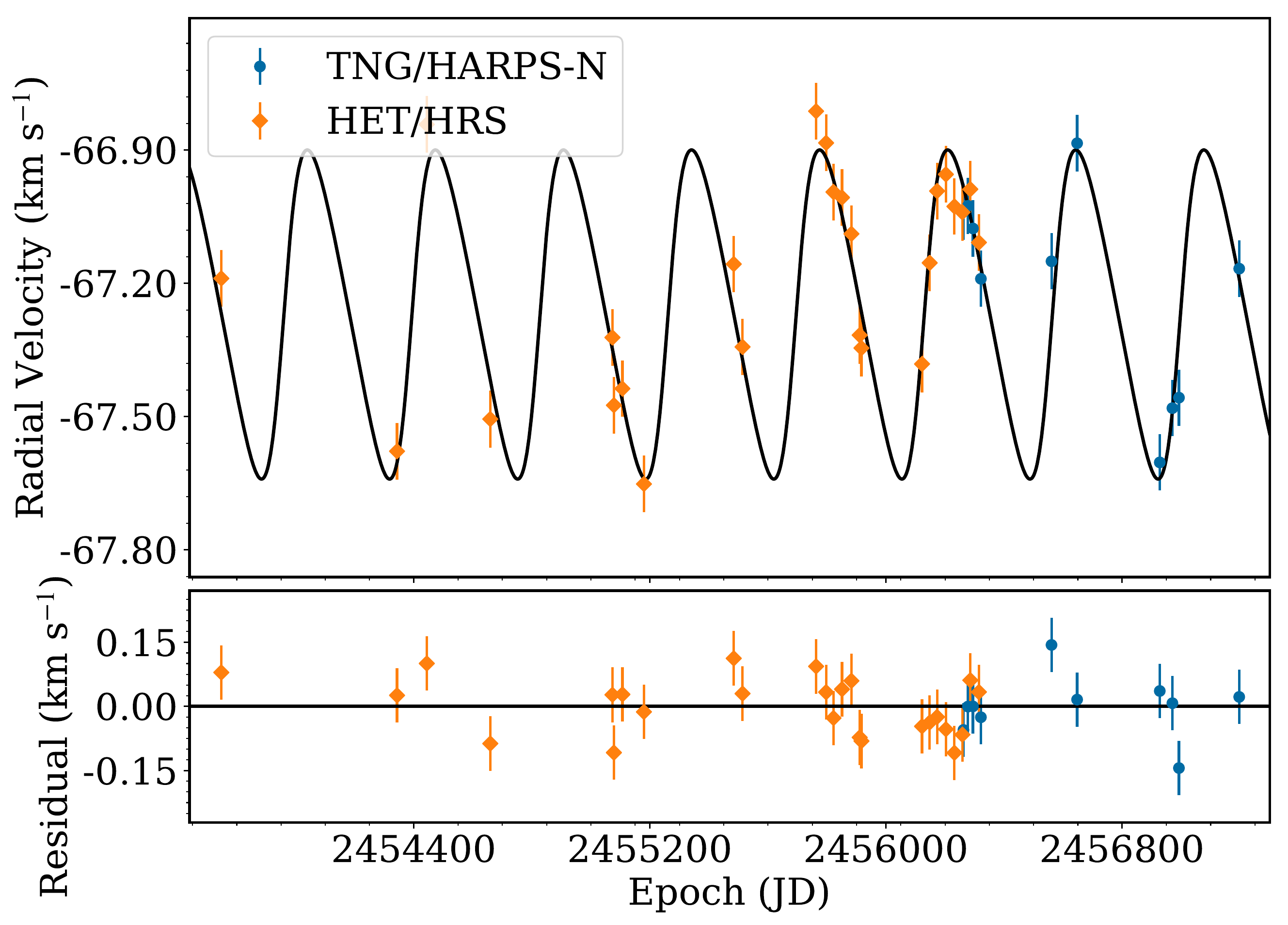}
   \caption{Keplerian best fit to combined HET/HRS (orange) and TNG/HARPS-N (blue) data for
      \starA. The jitter is added to uncertainties. The RV data show no amplitude or phase shift over 14 years of observations.}
   \label{Fit_0015_all}
\end{figure}

The high luminosity of HD 4760 makes it an interesting candidate for an LSP object. However, the existing photometric data from ASAS do not indicate any variability.
Moreover, our RV data covering about nine periods of the observed variation timescale, although not very numerous, do not show changes in amplitude or phase, as those detected in $\gamma$ Dra or $\alpha$ Tau  (Figure \ref{Fit_0015_all}).

The rotation period of HD 4760 (1531 days) 
is highly uncertain, and, given the uncertainties in $v\sin i$ and $\Rsun$, its maximum value ($P_{\mathrm{rot}}/\sin i$) may range between 672 and 8707 days. 
The orbital period from the Keplerian fit  to the RV data is shorter than the maximum allowed rotation period and we cannot exclude 
the possibility that the periodic distortions of spectral lines by a spot rotating with the star are the reason for the observed RV variations. 
However, HD 4760 does not show an increased activity (relative to the other stars in our sample) and none of activity indicators studied here is correlatec with the observed RV variations.
Also, an estimate of the spot fraction that would cause the observed RV amplitude, based on the scaling relation of \cite{Hatzes2002}, gives a rather unrealistic value of f=53[$\%$].
Our data also show that the  periodic RV variation have been present in HD 4760 for over nine years, which is unlikely, if caused by a surface feature.
Together with the apparent lack of photometric variability, we find that available data exclude that scenario.

We conclude that  the reflect motion due to a companion appears to be the most likely hypothesis that explains the observed RV variations in HD 4760. 

The mass of the companion and a rather tight orbit of HD 4760 b  locate it deep in the zone of engulfment \citep{VillaverLivio2009, Villaver2014, 2016RSOS....350571V}. 
However, predicting its future requires more detailed analysis, as this relatively massive companion may survive the common envelope phase of this system's evolution \citep{1984MNRAS.208..763L}.

See Table \ref{TableKeplerian} for details of the Keplerian model.

\subsection{HD 96992 } 

\begin{figure}
   \centering
   \includegraphics[width=0.5\textwidth]{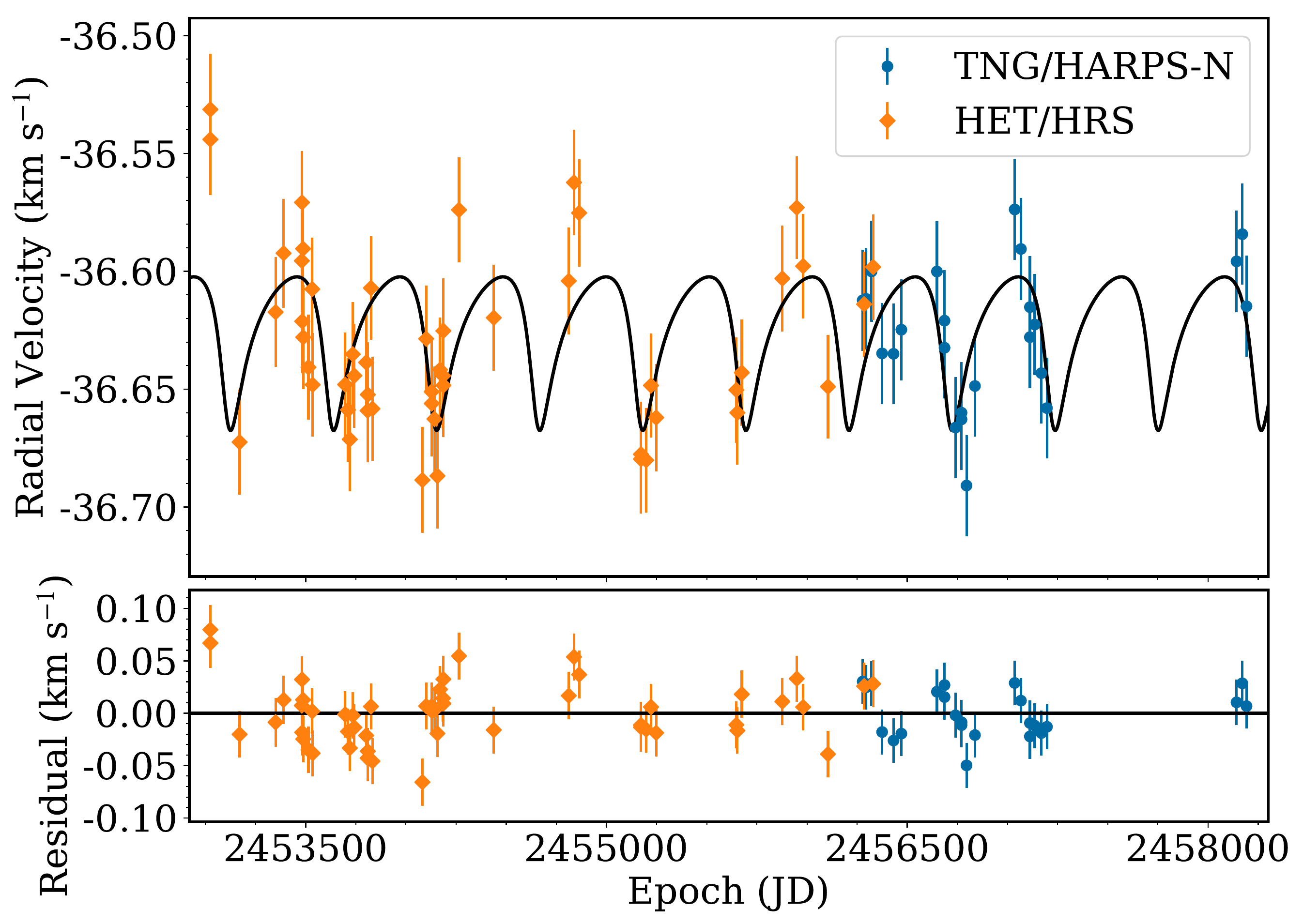}
   \caption{Keplerian best fit to combined HET/HRS (orange) and TNG/HARPS-N (blue) data for
      \starE. The jitter is added to uncertainties. }
   \label{Fit_0864_all}
\end{figure}

Of the time series presented in this paper, this is certainly the noisiest one. 
The Keplerian model for the 514-day period results in a RV semi-amplitude of only $33\ms$ (about five times greater than estimated HET/HRS precision), similar to the jitter of $20\ms$ (Figure \ref{Fit_0864_all}). 
The observed RV amplitude is about twenty  times larger than the expected amplitude of the p-mode oscillations. The jitter resulting from the Keplerian fit is larger than that expected from the scaling relations of \cite{KjeldsenBedding1995}.  This suggests an additional contribution, like granulation ,,flicker", to the jitter.

HD 96992 is much less luminous than HD 4760, with $\llsun$=1.47$\pm$0.09. It is located much below the TFGB, which makes it unlikely to be an irregular LSP giant. 
An apparent lack of photometric variability supports that claim as well. 

The orbital period of 514 days is much longer than the estimated rotation period of $198\pm92$ days ($P_{\mathrm{rot}}/\sin i=128-332$ days within uncertainties),
which, together with absence of a photometric variability of similar period and no correlation with activity indicators, excludes a spot rotating with the star as a cause of the observed RV variations.
The $\approx300$ days period present in RV residua is more likely to be due to rotation.

The apparent absence  of any correlation of observed RV variations with activity indicators and no trace of periodic variations in those indices makes the keplerian model the most consistent with the existing data.
Details of our Keplerian model are shown in  Table \ref{TableKeplerian}.

 The $m\sin i=1.14\pm0.31\Mjup$ planet of this system orbits the star deep in the engulfment zone \citep{VillaverLivio2009} and will most certainly be destroyed by its host before the AGB phase.


\subsection{BD+02 3313 } 

BD+02 3313 is a very intriguing case of a solar metallicity giant.
With $\llsun$=1.44$\pm$0.24 it is located well below the TFGB, even below the horizontal branch, which makes it very unlikely to be an LSP pulsating red giant.

The RV signal is very apparent;  the Keplerian orbit suggests a RV semi-amplitude of $690\ms$ and a period of 1393 days.

These RV data show an amplitude  over two orders of magnitude larger than that expected of the p-modes oscillations. 

The fitted jitter of 10 ms$^{-1}$ is close to the  expected from the scaling relations of \cite{KjeldsenBedding1995}, within uncertainties of mass and luminosity. 

The estimated rotation period of $238\pm122$ days ($P_{\mathrm{rot}}/\sin i=146-421$ within uncertainties)  is much shorter than the Keplerian orbital period. 
The extensive photometric data set from ASAS, contemporaneous with our HET/HRS data, shows no periodic signal
and no excess scatter that might be a signature of spots on the surface of the star. 

None of the  activity indices studied here shows a significant periodic signal. Line bisectors, $\ha$ and $\shk$ are uncorrelated with the RV variations. The value of $\shk$ does not indicate a significant activity, compared to other stars in our sample.
The persistence of the RV periodicity for over 11 years also advocates against a possible influence of an active region rotating with the star.

The resulting Keplerian model (Figure \ref{Fit_0128_all}), which suggests an $m\sin i=34.1\pm1.1\Mjup$ companion in a 2.47 au, eccentric ($e=0.47$)
orbit (i.e., a brown dwarf in the brown dwarf desert)  is consistent with the available RV data for the total time-span of the observing run.

However, 
FWHM of the CCF from HARPS-N data for BD+02~3313  shows an $\mathrm{r}=0.73 >\mathrm{r}_{c}=0.62$ correlation, which is statistically significant  at the accepted confidence level of p=0.01 (Figure \ref{rv_fwhm_fig}, lower left panel). 
Given the small number of CCF FWHM data points, we cannot exclude the possibility that the observed correlation is spurious.
This possibility seems to be supported by the apparent lack of a periodic signal in the LS periodogram for CCF FWHM (Figure \ref{LSP_0128}).

\begin{figure}
   \centering
   \includegraphics[width=0.5\textwidth]{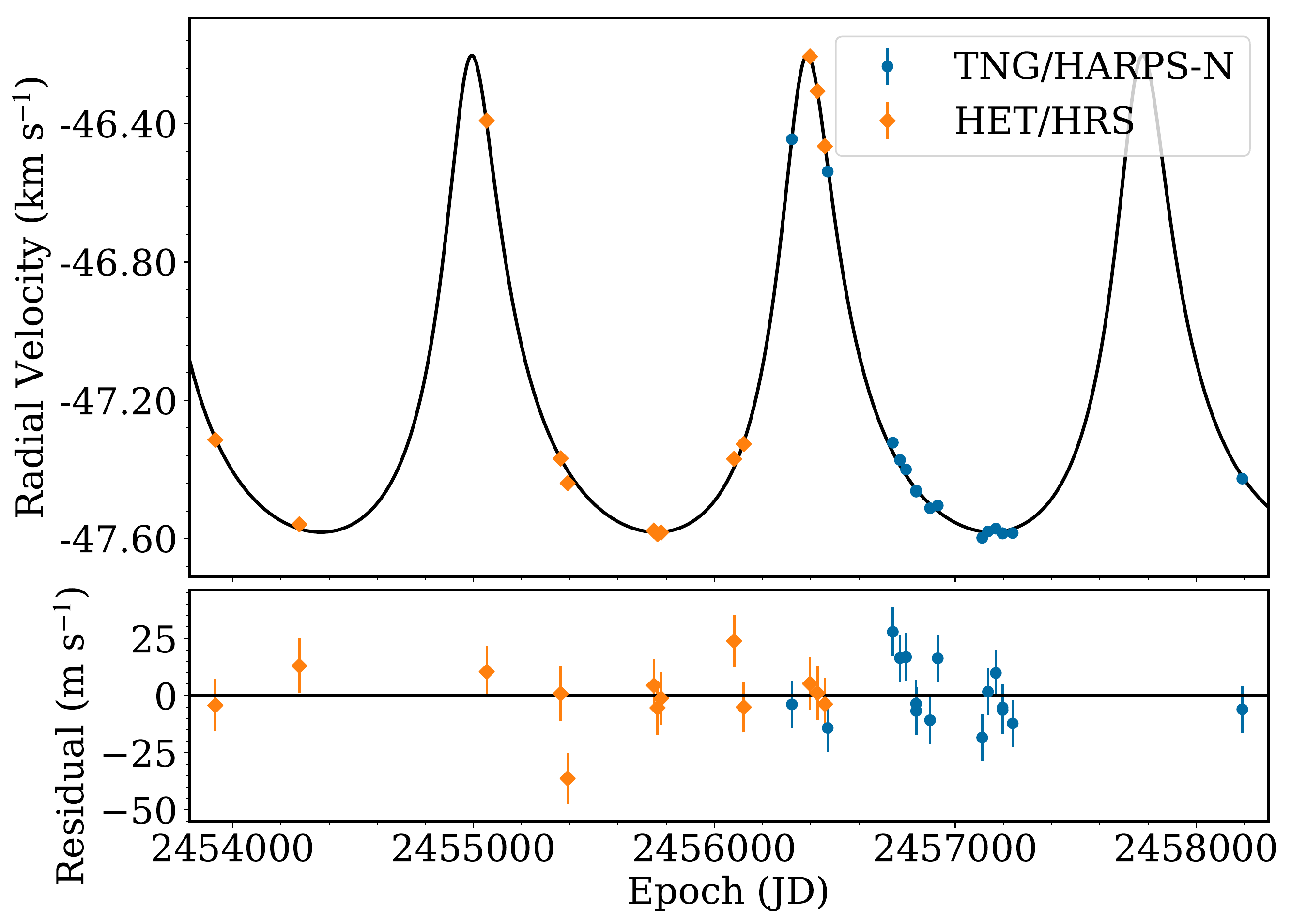}
   \caption{Keplerian best fit to combined HET/HRS (orange) and TNG/HARPS-N (blue) data for
      \starB. The jitter is added to uncertainties. The RV signal is very apparent.}
   \label{Fit_0128_all}
\end{figure} 

An assumption that all the observed RV variations in this inactive star are due to the rotation of a surface feature is inconsistent with the existing photometry and our rotation period estimate.
A more likely explanation would be the presence of a spatially unresolved companion associated with  BD+02 3313. 

We conclude that the observed RV and CCF FWHM correlation seriously undermines 
the Keplerian model of the observed RV variations in BD+02~3313. The actual cause of the reported RV variations remains to be identified with the help of additional observations.

 \subsection{TYC 0434-04538-1 } 

 TYC 0434-04538-1 is a low metallicity, [Fe/H]=-0.38$\pm$0.06 giant, with a luminosity of $\llsun=1.67\pm0.09$, which locates it near the horizontal branch.
 As such, the star is unlikely to be an irregular LSP giant. 
 
 It shows a strong, periodic RV signal, which, when modelled under the assumption of a Keplerian motion, shows a semi amplitude of $K=209\ms$, and a period of 193 days. 
 The RV data show an amplitude   about forty times larger than that expected of the p-mode oscillation.  Again, the jitter is larger than expected form the p-mode oscillations only, so it likely contains an additional component, unresolved by our observations, like the granulation ,,flicker".
 
 This period is shorter than the estimate of $P_{\mathrm{rot}}/\sin i=124-225$~days, hence the observed RV variation may originate, in principle,
 from a feature on the stellar surface rotating with the star. 
 The spot would have to cover f=$10\%$ of the stellar surface according to the simple model of Hatzes  \cite{Hatzes2002} to explain the observed RV variations. 
 Photometric data from ASAS, which show no variability, do not support this scenario.
 We also note that such a large spot coverage ($10\%$) was successfully applied to model spots on the surface of the overactive spotted giant in a binary system EPIC 211759736 by \cite{2018A&A...620A.189O}.
 
 Consequently, we conclude that the available data favour the low-mass companion hypothesis. 
 
\begin{figure}
   \centering
   \includegraphics[width=0.5\textwidth]{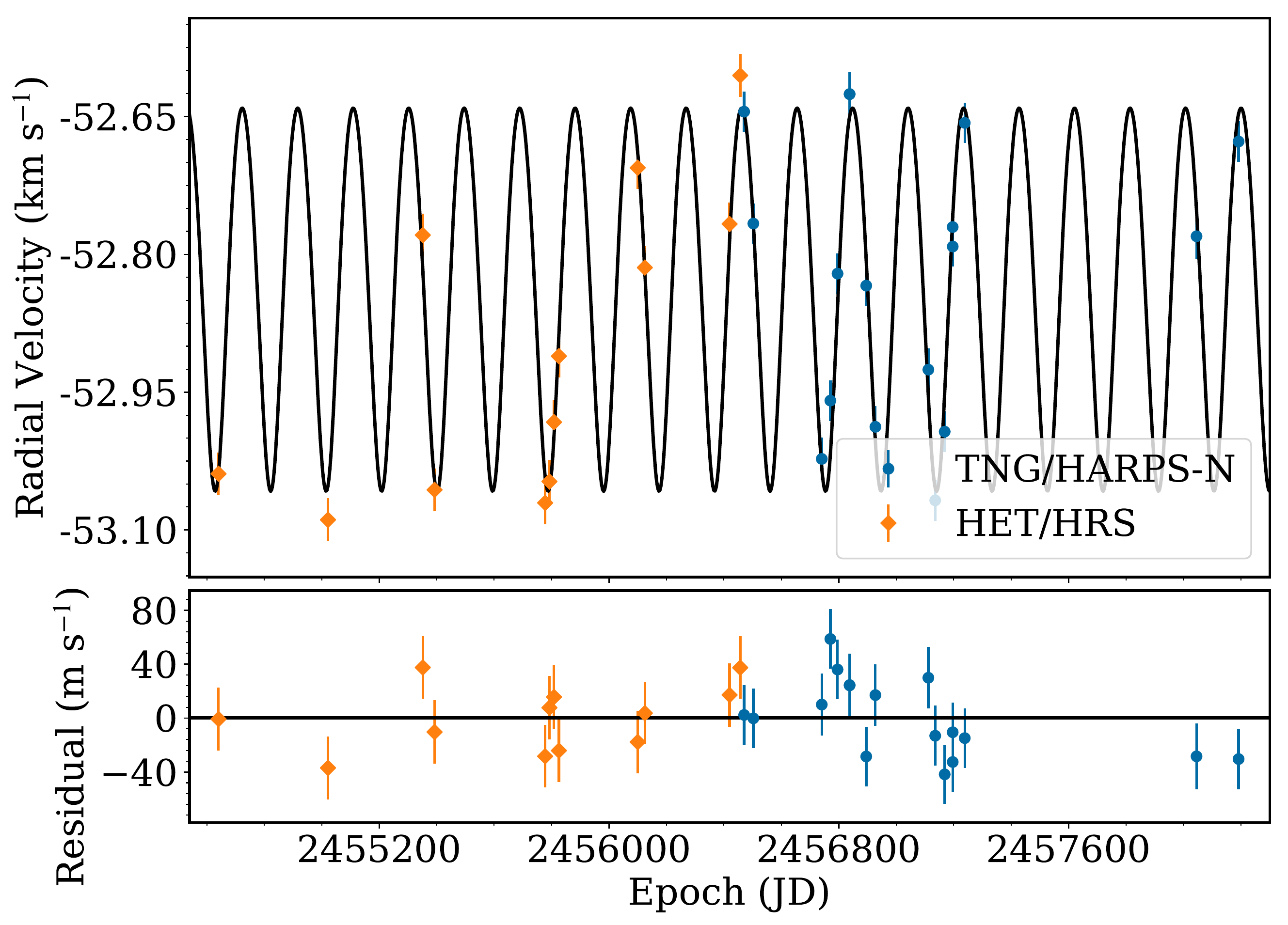}
   \caption{Keplerian best fit to combined HET/HRS (orange) and TNG/HARPS-N (blue) data for
      \starC. The jitter is added to uncertainties.  The RV variations appear to be stable over the period of nearly 10 years.}
   \label{Fit_0128_all}
\end{figure}

\begin{figure}
   \centering
   \includegraphics[width=0.5\textwidth]{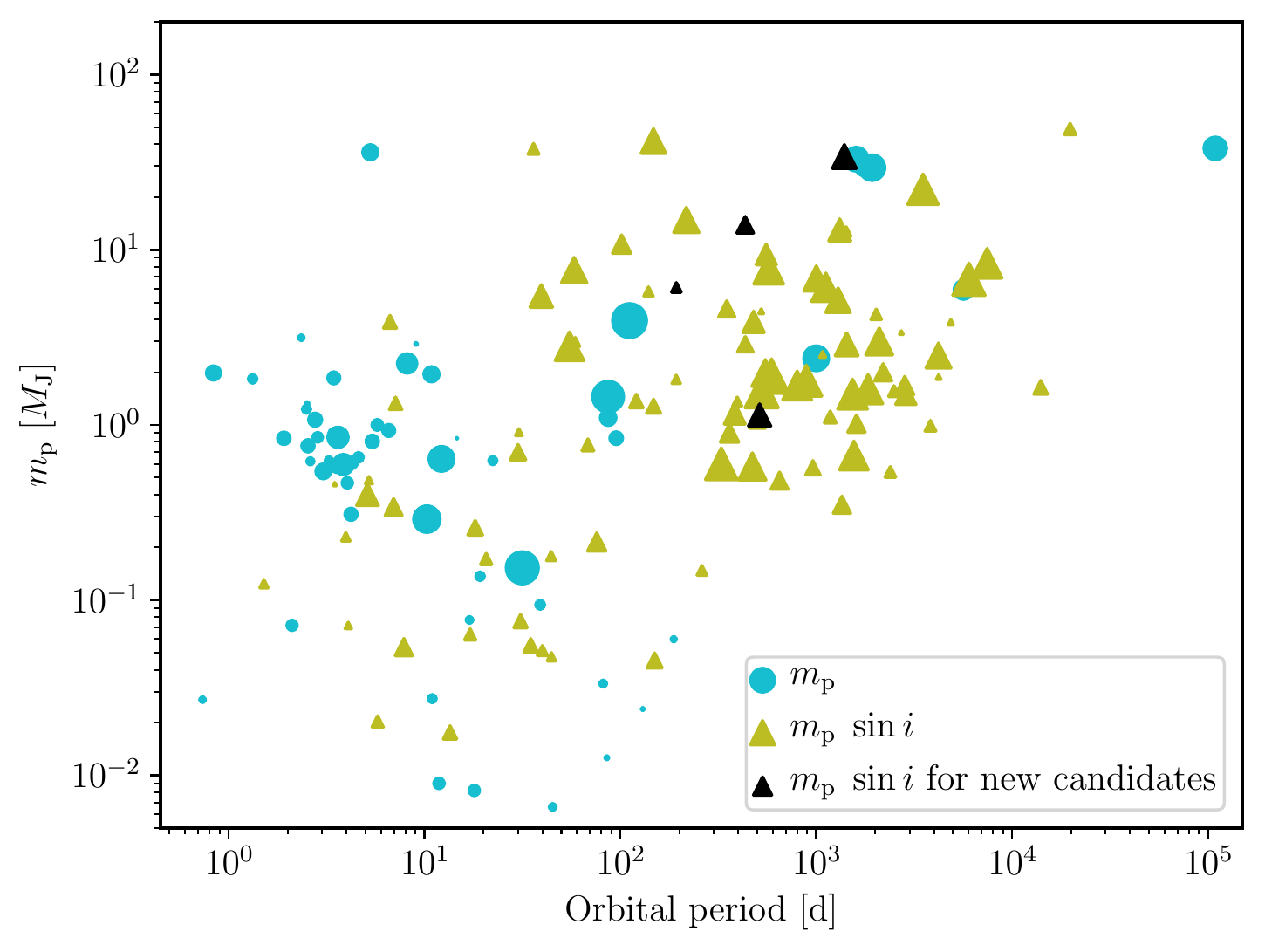}
   \caption{Mass-orbital period relation for 228 planets hosted  by solar mass stars (within 5$\%$) in exoplanets.org, together with our four new candidates presented here. Symbol sizes are scaled with orbital eccentricities. }
   \label{mass-period}
\end{figure}

\subsection{The current status of the project}

{The 
sample contains 122  stars in total, with at least two epochs 
of observations  that allowed us to measure the RV variation amplitude.}

 Sixty stars in the sample (49$\pm5\%$) are assumed to be single, as they show $\Delta\mathrm{RV}<50\ms$
 in several epochs of data over a period of, typically, 2-3 years.
 This group of stars 
 may still contain  long period and/or low-mass companions, which means that the number of single stars may be overestimated. 
 { Due to available
telescope time further observations of these stars have been ceased. }

The estimate of single star frequency in our sample, although based on a small sample of GK stars
at various stages of evolution from the MS to the RGB, agrees with 
the results of a  study of a sample of 454 stars by \citealt{2010ApJS..190....1R} who found that $54\pm 2\%$  of solar-type stars are single.
We take this agreement as a confirmation that our sample is not biased towards binary or single stars.

Nineteen stars in our sample ($16\pm3\%$) are spectroscopic binaries with $\Delta\mathrm{RV}>2\kms$.
Technically not only  HD 181368 b \citep{2018A&A...613A..47A} but also BD+20 274 b \citep{Gettel2012b}
 belongs to this group, due to the observed RV trend. Although we cannot exclude more low-mass companions associated with binary stellar systems for these targets, 
 they were rejected from further observations after several epochs, due to a limited telescope time available.

 
 Finally, 43 of the stars in our sample ($35\pm4\%$) show RV amplitudes between $50\ms$ and $2\kms$ and are assumed to be either active stars or 
 planetary/BD companion candidates. These stars have been searched for low-mass companions by this project. 
 
  Six low-mass companion hosts have been identified in the  sample so far: 
 HD 102272 \citep{Niedzielski2009a}, 
 BD+20 2457 \citep{Niedzielski2009b}, 
 BD+20 274 \citep{Gettel2012b}, 
 HD 219415 \citep{Gettel2012b}, 
 HD 5583 \citep{Niedzielski2016b}, 
 and HD 181368 \citep{2018A&A...613A..47A}.

This paper presents  low-mass companions to another three stars: 
HD 4760,  
TYC 0434 04538 1, 
HD 96992. 
Our findings add to a population of 228 planets orbiting solar-mass stars in exoplanets.org (Figure \ref{mass-period}).

Seven stars from the sample 
(HD 102272, BD+20 2457,   BD+20 274,  HD 219415,  HD 5583, 
TYC 0434 04538 1, HD 96992)
show RV variations consistent with planetary-mass companions ($m_\mathrm{p} \sin i < 13 \Mjup$), which represents $6\pm2\%$ of the sample.

 \section{Conclusions}
 
Based on precise RV measurements gathered with the HET/HRS and Harps-N for over 11 years, we have discussed three solar-mass giants with low mass companions: 
 HD 4760 hosts a $m\sin i=13.9\pm 2.4\Mjup$ companion in an $a=1.14\pm0.08$~au and $e=0.23\pm0.09$ orbit;
 HD 96992 has a $m\sin i=1.14\pm 0.31\Mjup$ companion in an $a=1.24\pm0.05$~au, eccentric, $e=0.41\pm0.24$ orbit;
 TYC 0434-04538-1 is accompanied with a $m \sin i=6.1\pm 0.7\Mjup$ companion in an $a=1.66\pm0.04$~au, nearly circular  orbit with $e=0.08\pm0.05$.
 In the case of BD+02 3313 we find the Keplerian model uncertain because of statistically significant correlation between RV and CCF FWHM in the HARPS-N data.
 
 The analysis of RV amplitudes in our sample of 122 solar-mass stars at various stellar evolution stages shows that single star frequency is  $49\pm5\%$, which means that the sample is not biased against stellar binarity.
 
\begin{acknowledgements}
 
We thank the HET  and IAC resident astronomers and telescope operators  for
their support.

AN  was supported by the Polish National Science Centre
grant no. 2015/19/B/ST9/02937.

EV acknowledges support from the Spanish Ministerio de Ciencia Inovación y Universidades  under grant PGC2018-101950-B-100. 
KK was funded in part by the Gordon and Betty Moore Foundation's
Data-Driven Discovery Initiative through Grant GBMF4561.

This research was supported in part by PL-Grid Infrastructure.

The HET is a
joint project of the University of Texas at Austin, the Pennsylvania State
University, Stanford University, Ludwig- Maximilians-Universit\"at M\"unchen,
and Georg-August-Universit\"at G\"ottingen. The HET is named in honor of its
principal benefactors, William P. Hobby and Robert E. Eberly. 
The Center for Exoplanets and Habitable Worlds is supported by the Pennsylvania State
University, the Eberly College of Science, and the Pennsylvania Space Grant
Consortium.

This research has made use of the SIMBAD database, operated at CDS, Strasbourg, France. This research has made use of
NASA's Astrophysics Data System. The acknowledgements were compiled using the Astronomy Acknowledgement Generator.
This research made use of SciPy~\citep{Scipy2001}. This research made use of the yt-project, a toolkit for
analyzing and visualizing quantitative data~\citep{yt}. This research made use of matplotlib, a Python library
for publication quality graphics~\citep{mpl}. This research made use of Astropy, a community-developed core
Python package for Astronomy~\citep{Astropy2013}.  IRAF is distributed by the National Optical Astronomy
Observatory, which is operated by the Association of Universities for Research in Astronomy (AURA) under cooperative
agreement with the National Science Foundation~\citep{Iraf1993}. This research made use of
NumPy~\citep{numpy}. 

We thank the referee for comments that have significantly contributed to improving this paper.

\end{acknowledgements}
\bibliographystyle{aa} 
\bibliography{an2} 
\end{document}

%% file: table_RV_BIS.tex
\begin{longtable}{lcrrrrr}  

\caption{RV data for program stars. Available electronically only.}\\ 
\hline
Star & Instrument & JD & RV & \srv & BIS  & \sbs \\ 
& &  & [\ms] & [\ms] & [\ms] &  [\ms] \\\hline\hline 
 \endfirsthead\caption{continued}\\ 
 \hline 
Star & Instrument & JD & RV & \srv & BIS  & \sbs \\ 
& &  & [\ms] & [\ms] & [\ms] &  [\ms] \\ 
 \hline \hline 
\endhead
         HD 4760 & HET &  2453747.575694 &   -25.86 &   5.99 &  67.43 &  18.27 \\
         HD 4760 & HET &  2454342.946962 &  -414.81 &   7.76 &  79.03 &  20.63 \\
         HD 4760 & HET &  2454443.681389 &   320.76 &   6.23 &  72.93 &  19.24 \\
         HD 4760 & HET &  2454658.950532 &  -342.45 &   9.33 &  23.85 &  35.21 \\
         HD 4760 & HET &  2455072.807951 &  -158.79 &   9.75 & 144.67 &  11.45 \\
         HD 4760 & HET &  2455077.796435 &  -311.32 &   6.28 &  75.36 &  11.89 \\
         HD 4760 & HET &  2455106.720556 &  -273.87 &   6.54 &  67.99 &  12.25 \\
         HD 4760 & HET &  2455179.534306 &  -488.58 &   6.10 &  30.22 &  14.37 \\
         HD 4760 & HET &  2455483.837199 &     6.40 &   6.33 &  79.23 &  15.13 \\
         HD 4760 & HET &  2455513.741308 &  -180.02 &   5.87 &  26.91 &  16.29 \\
         HD 4760 & HET &  2455762.902917 &   350.44 &   7.07 &  23.70 &  22.83 \\
         HD 4760 & HET &  2455796.830660 &   279.09 &   6.87 &  68.37 &  15.65 \\
         HD 4760 & HET &  2455821.906296 &   168.59 &   7.07 &  91.88 &  21.23 \\
         HD 4760 & HET &  2455850.831181 &   156.20 &   5.16 & 121.96 &  13.70 \\
         HD 4760 & HET &  2455882.597899 &    74.40 &   5.52 &  45.03 &  14.47 \\
         HD 4760 & HET &  2455910.652147 &  -153.54 &  10.28 &  39.68 &  42.27 \\
         HD 4760 & HET &  2455916.658819 &  -182.66 &   5.98 &  79.49 &  16.06 \\
         HD 4760 & HET &  2456121.948206 &  -218.47 &   6.75 &  66.14 &  14.04 \\
         HD 4760 & HET &  2456147.864815 &     9.10 &   6.12 &  75.62 &  11.69 \\
         HD 4760 & HET &  2456172.954363 &   170.80 &   6.66 &   8.82 &  16.76 \\
         HD 4760 & HET &  2456202.726285 &   208.31 &   6.06 &  81.47 &  15.59 \\
         HD 4760 & HET &  2456231.634751 &   136.16 &   5.29 & 117.31 &  14.07 \\
         HD 4760 & HET &  2456258.711968 &   122.91 &   6.30 & 113.09 &  18.08 \\
         HD 4760 & HET &  2456285.641852 &   174.83 &   5.87 &  96.21 &  15.51 \\
         HD 4760 & HET &  2456314.572708 &    54.92 &   6.28 & 145.80 &  18.11 \\
         HD 4760 & TNG & 2456262.357503 & -67038.55 &   0.94 & 209.20 &  \\
         HD 4760 & TNG & 2456277.467716 & -67025.63 &   1.31 & 195.75 &  \\
         HD 4760 & TNG & 2456294.439304 & -67076.55 &   0.79 & 216.56 &  \\
         HD 4760 & TNG & 2456321.387544 & -67189.82 &   0.83 & 242.55 &  \\
         HD 4760 & TNG & 2456560.680562 & -67150.42 &   1.05 & 155.16 &  \\
         HD 4760 & TNG & 2456647.450312 & -66884.74 &   1.08 & 252.03 &  \\
         HD 4760 & TNG & 2456927.561058 & -67603.04 &   1.26 & 176.64 &  \\
         HD 4760 & TNG & 2456970.334628 & -67480.96 &   0.91 & 162.39 &  \\
         HD 4760 & TNG & 2456992.493091 & -67457.37 &   1.44 & 190.77 &  \\
         HD 4760 & TNG & 2457196.727447 & -67167.02 &   1.34 & 170.63 &  \\
        HD 96992 & HET &  2453024.829317 &    85.98 &   9.78 &   7.21 &  22.13 \\
        HD 96992 & HET &  2453024.835185 &    98.72 &  10.02 &   9.17 &  24.41 \\
        HD 96992 & HET &  2453170.640087 &   -42.42 &   6.23 &  59.20 &  10.37 \\
        HD 96992 & HET &  2453349.927853 &    12.73 &   9.24 & -39.90 &  21.44 \\
        HD 96992 & HET &  2453389.802054 &    37.71 &   8.79 &  33.37 &   9.71 \\
        HD 96992 & HET &  2453480.780069 &    34.49 &   4.24 &  15.12 &  10.89 \\
        HD 96992 & HET &  2453481.779132 &    59.27 &   4.77 &  26.70 &  10.20 \\
        HD 96992 & HET &  2453482.773252 &     8.81 &   5.02 &   6.68 &  10.38 \\
        HD 96992 & HET &  2453486.759965 &    39.63 &   4.89 &  31.00 &  13.91 \\
        HD 96992 & HET &  2453487.760810 &     2.07 &   5.28 &  21.80 &  14.40 \\
        HD 96992 & HET &  2453513.690370 &   -10.68 &   6.28 &  -8.56 &  10.84 \\
        HD 96992 & HET &  2453532.647419 &    22.48 &   4.76 &  17.77 &  10.46 \\
        HD 96992 & HET &  2453534.657234 &   -18.09 &   4.68 &   9.77 &  10.72 \\
        HD 96992 & HET &  2453696.951944 &   -18.01 &   5.53 & -17.28 &  14.21 \\
        HD 96992 & HET &  2453710.923738 &   -28.79 &   5.38 & -32.71 &  13.96 \\
        HD 96992 & HET &  2453719.892535 &   -41.26 &   5.65 &  37.70 &  14.98 \\
        HD 96992 & HET &  2453734.848866 &    -5.15 &   5.85 &  -7.57 &  17.45 \\
        HD 96992 & HET &  2453742.827083 &   -14.22 &   5.43 &  28.32 &  13.82 \\
        HD 96992 & HET &  2453801.690660 &    -8.65 &   6.52 &  18.57 &  14.05 \\
        HD 96992 & HET &  2453808.644375 &   -29.06 &   4.99 &  18.69 &  13.09 \\
        HD 96992 & HET &  2453809.642627 &   -22.27 &   5.86 &  25.81 &  16.90 \\
        HD 96992 & HET &  2453825.610498 &    22.97 &   4.83 &  15.16 &  11.39 \\
        HD 96992 & HET &  2453832.823229 &   -28.26 &   5.17 &  29.39 &  12.17 \\
        HD 96992 & HET &  2454081.897593 &   -58.52 &   7.03 &  32.81 &  17.34 \\
        HD 96992 & HET &  2454100.853692 &     1.43 &   7.02 &  48.75 &  21.19 \\
        HD 96992 & HET &  2454127.789873 &   -21.05 &   7.18 &  26.00 &  17.38 \\
        HD 96992 & HET &  2454128.783588 &   -26.05 &   6.49 &  27.38 &  15.04 \\
        HD 96992 & HET &  2454141.990775 &   -32.62 &   6.11 & -18.10 &  13.00 \\
        HD 96992 & HET &  2454156.942813 &   -56.78 &   6.67 &  36.54 &  16.87 \\
        HD 96992 & HET &  2454169.893299 &   -11.73 &   5.68 & -23.33 &  17.66 \\
        HD 96992 & HET &  2454184.635822 &   -14.25 &   6.12 &  29.08 &  21.51 \\
        HD 96992 & HET &  2454186.620301 &     4.79 &   6.40 &   1.21 &  18.99 \\
        HD 96992 & HET &  2454186.871007 &   -18.34 &   4.89 &  14.12 &  13.32 \\
        HD 96992 & HET &  2454264.639271 &    56.10 &   6.34 & -20.46 &  14.63 \\
        HD 96992 & HET &  2454437.932813 &    10.33 &   6.79 & -29.68 &  14.89 \\
        HD 96992 & HET &  2454811.922894 &    25.98 &   7.43 & -19.93 &  22.46 \\
        HD 96992 & HET &  2454837.846481 &    67.73 &   6.48 &  52.62 &  16.37 \\
        HD 96992 & HET &  2454863.768970 &    54.81 &   7.83 &  46.48 &  22.81 \\
        HD 96992 & HET &  2455171.927662 &   -49.61 &   9.05 &  -3.87 &  26.08 \\
        HD 96992 & HET &  2455171.935185 &   -47.60 &   6.54 &  -8.78 &  20.62 \\
        HD 96992 & HET &  2455197.855943 &   -50.09 &   5.71 &   8.33 &  15.82 \\
        HD 96992 & HET &  2455222.017373 &   -18.40 &   5.23 &  36.74 &  16.57 \\
        HD 96992 & HET &  2455247.712338 &   -32.02 &   7.93 &   9.62 &  25.26 \\
        HD 96992 & HET &  2455647.625150 &   -20.32 &   6.49 &  14.47 &  15.44 \\
        HD 96992 & HET &  2455652.610532 &   -29.97 &   5.60 &  32.83 &  12.39 \\
        HD 96992 & HET &  2455674.782118 &   -12.95 &   7.45 &  45.98 &  14.90 \\
        HD 96992 & HET &  2455877.003889 &    26.99 &   6.86 &  -0.06 &  18.67 \\
        HD 96992 & HET &  2455948.799039 &    57.06 &   4.45 &  35.93 &  12.12 \\
        HD 96992 & HET &  2455980.721296 &    32.25 &   5.71 &  33.10 &  13.58 \\
        HD 96992 & HET &  2456104.620208 &   -18.90 &   5.32 & -42.61 &  12.09 \\
        HD 96992 & HET &  2456284.872350 &    16.08 &   6.37 &  33.79 &  17.88 \\
        HD 96992 & HET &  2456329.740498 &    31.88 &   6.65 &  10.72 &  18.51 \\
        HD 96992 & TNG & 2456277.741075 & -36612.30 &   1.26 &  80.20 &  \\
        HD 96992 & TNG & 2456294.661935 & -36611.70 &   0.88 &  89.68 &  \\
        HD 96992 & TNG & 2456321.656079 & -36600.06 &   1.14 &  79.65 &  \\
        HD 96992 & TNG & 2456374.540529 & -36634.85 &   1.40 &  93.80 &  \\
        HD 96992 & TNG & 2456431.433702 & -36635.03 &   1.01 &  89.87 &  \\
        HD 96992 & TNG & 2456470.371400 & -36624.79 &   1.57 &  85.34 &  \\
        HD 96992 & TNG & 2456647.720323 & -36600.10 &   1.49 &  87.07 &  \\
        HD 96992 & TNG & 2456685.652759 & -36620.97 &   1.87 &  93.25 &  \\
        HD 96992 & TNG & 2456685.708909 & -36632.45 &   1.78 &  85.68 &  \\
        HD 96992 & TNG & 2456740.529729 & -36666.30 &   1.46 &  84.68 &  \\
        HD 96992 & TNG & 2456770.497979 & -36662.81 &   1.25 &  83.07 &  \\
        HD 96992 & TNG & 2456770.535865 & -36660.03 &   1.14 &  85.80 &  \\
        HD 96992 & TNG & 2456795.441384 & -36690.91 &   1.50 &  95.74 &  \\
        HD 96992 & TNG & 2456837.383901 & -36648.67 &   1.48 &  88.55 &  \\
        HD 96992 & TNG & 2457035.643601 & -36573.76 &   1.52 &  83.60 &  \\
        HD 96992 & TNG & 2457066.696030 & -36590.54 &   3.23 &  77.73 &  \\
        HD 96992 & TNG & 2457111.392729 & -36627.95 &   2.64 &  77.58 &  \\
        HD 96992 & TNG & 2457111.558431 & -36615.20 &   3.90 &  94.12 &  \\
        HD 96992 & TNG & 2457135.574680 & -36622.59 &   1.37 &  81.68 &  \\
        HD 96992 & TNG & 2457168.434249 & -36643.22 &   1.09 &  91.15 &  \\
        HD 96992 & TNG & 2457196.410659 & -36658.05 &   1.07 &  91.97 &  \\
        HD 96992 & TNG & 2458141.769519 & -36595.78 &   2.58 & 101.38 &  \\
        HD 96992 & TNG & 2458170.647338 & -36584.26 &   1.55 &  79.80 &  \\
        HD 96992 & TNG & 2458191.578169 & -36614.83 &   1.34 &  96.03 &  \\
      BD+02 3313 & HET &  2453927.778171 &  -208.64 &   5.07 &   4.09 &  11.53 \\
      BD+02 3313 & HET &  2454276.705550 &  -452.92 &   6.14 & -25.92 &  15.72 \\
      BD+02 3313 & HET &  2455054.693333 &   714.90 &   4.91 & -37.46 &   8.01 \\
      BD+02 3313 & HET &  2455361.848137 &  -262.40 &   6.33 &  13.39 &  16.11 \\
      BD+02 3313 & HET &  2455390.646667 &  -334.08 &   4.64 &  28.05 &  10.25 \\
      BD+02 3313 & HET &  2455748.668484 &  -471.04 &   5.45 & -11.68 &  11.26 \\
      BD+02 3313 & HET &  2455763.736586 &  -480.98 &   5.89 & -26.64 &  13.12 \\
      BD+02 3313 & HET &  2455778.716956 &  -476.43 &   5.48 & -33.29 &  12.86 \\
      BD+02 3313 & HET &  2456081.878576 &  -263.56 &   5.17 &  47.84 &  15.96 \\
      BD+02 3313 & HET &  2456121.765046 &  -220.38 &   4.06 & -29.51 &  10.42 \\
      BD+02 3313 & HET &  2456396.889479 &   900.27 &   5.51 & -15.85 &  14.16 \\
      BD+02 3313 & HET &  2456427.940463 &   800.11 &   5.45 &  -8.91 &  14.95 \\
      BD+02 3313 & HET &  2456458.850289 &   640.52 &   5.09 & -16.52 &  12.06 \\
      BD+02 3313 & TNG & 2456321.797887 & -46444.51 &   0.78 &  64.27 &  \\
      BD+02 3313 & TNG & 2456470.496536 & -46538.09 &   1.47 &  66.68 &  \\
      BD+02 3313 & TNG & 2456740.664455 & -47322.35 &   2.71 &  83.27 &  \\
      BD+02 3313 & TNG & 2456770.628872 & -47372.23 &   0.96 &  73.27 &  \\
      BD+02 3313 & TNG & 2456795.572893 & -47399.84 &   2.32 &  69.10 &  \\
      BD+02 3313 & TNG & 2456837.424829 & -47460.48 &   1.47 &  77.02 &  \\
      BD+02 3313 & TNG & 2456837.628672 & -47463.79 &   2.26 &  58.28 &  \\
      BD+02 3313 & TNG & 2456895.478523 & -47511.77 &   1.34 &  73.39 &  \\
      BD+02 3313 & TNG & 2456927.361856 & -47504.02 &   1.97 &  74.65 &  \\
      BD+02 3313 & TNG & 2457111.733353 & -47597.61 &   1.51 &  75.04 &  \\
      BD+02 3313 & TNG & 2457135.655325 & -47579.18 &   1.60 &  71.61 &  \\
      BD+02 3313 & TNG & 2457168.545205 & -47571.03 &   0.77 &  74.59 &  \\
      BD+02 3313 & TNG & 2457196.440339 & -47584.06 &   1.29 &  75.88 &  \\
      BD+02 3313 & TNG & 2457196.576370 & -47585.12 &   1.40 &  75.61 &  \\
      BD+02 3313 & TNG & 2457238.433643 & -47583.82 &   1.41 &  77.04 &  \\
      BD+02 3313 & TNG & 2458191.746301 & -47426.28 &   1.46 &  69.25 &  \\
TYC 0434-04538-1 & HET &  2454640.738762 &  -141.84 &   8.19 &  18.23 &  24.03 \\
TYC 0434-04538-1 & HET &  2455021.815810 &  -191.91 &   8.18 &  20.40 &  11.65 \\
TYC 0434-04538-1 & HET &  2455351.888883 &   117.97 &   7.54 & -42.20 &  19.22 \\
TYC 0434-04538-1 & HET &  2455392.786684 &  -159.35 &   8.39 &  -2.09 &  25.54 \\
TYC 0434-04538-1 & HET &  2455777.732801 &  -173.52 &   7.57 &  25.37 &  18.12 \\
TYC 0434-04538-1 & HET &  2455792.682598 &  -150.22 &   8.49 & -40.49 &  21.30 \\
TYC 0434-04538-1 & HET &  2455808.662020 &   -85.64 &   9.08 & -41.68 &  21.19 \\
TYC 0434-04538-1 & HET &  2455825.618692 &   -13.78 &   7.92 & -23.07 &  20.22 \\
TYC 0434-04538-1 & HET &  2456099.863345 &   191.25 &   7.32 &  43.01 &  20.40 \\
TYC 0434-04538-1 & HET &  2456124.799143 &    82.62 &   7.58 &  13.99 &  19.32 \\
TYC 0434-04538-1 & HET &  2456419.865567 &   130.05 &   8.64 &  33.65 &  21.10 \\
TYC 0434-04538-1 & HET &  2456456.873542 &   291.73 &   7.80 &  21.88 &  20.67 \\
TYC 0434-04538-1 & TNG & 2456470.520578 & -52644.69 &   1.89 & 105.35 &  \\
TYC 0434-04538-1 & TNG & 2456502.579924 & -52766.47 &   1.75 &  69.31 &  \\
TYC 0434-04538-1 & TNG & 2456740.685233 & -53022.67 &   7.39 &  87.55 &  \\
TYC 0434-04538-1 & TNG & 2456770.650312 & -52959.20 &   2.85 &  82.27 &  \\
TYC 0434-04538-1 & TNG & 2456795.601898 & -52820.97 &   2.88 &  78.46 &  \\
TYC 0434-04538-1 & TNG & 2456837.529009 & -52625.57 &   5.57 & 114.88 &  \\
TYC 0434-04538-1 & TNG & 2456837.649264 & -52625.57 &   8.99 & 103.96 &  \\
TYC 0434-04538-1 & TNG & 2456895.499521 & -52834.08 &   3.01 &  96.20 &  \\
TYC 0434-04538-1 & TNG & 2456927.400101 & -52987.73 &   5.93 &  70.48 &  \\
TYC 0434-04538-1 & TNG & 2457111.752827 & -52925.48 &   6.75 &  95.20 &  \\
TYC 0434-04538-1 & TNG & 2457135.673906 & -53067.80 &   4.27 &  88.03 &  \\
TYC 0434-04538-1 & TNG & 2457168.568493 & -52993.00 &   2.32 &  77.29 &  \\
TYC 0434-04538-1 & TNG & 2457196.445648 & -52770.23 &   2.84 & 102.66 &  \\
TYC 0434-04538-1 & TNG & 2457196.583515 & -52791.48 &   2.18 & 100.05 &  \\
TYC 0434-04538-1 & TNG & 2457238.453013 & -52656.91 &   2.85 & 112.99 &  \\
TYC 0434-04538-1 & TNG & 2458045.363404 & -52780.32 &  10.92 & 100.10 &  \\
TYC 0434-04538-1 & TNG & 2458191.753797 & -52677.18 &   4.60 &  67.24 &  \\
 \hline 
  \label{RV-DATA} 
 \end{longtable} 

%% file: table_master-new.tex
\renewcommand{\arraystretch}{1.3}
\begin{tabular}{llllll}
\hline  
  Parameter               &  \scriptsize  \starA   & \scriptsize \starB       & \scriptsize  \starC    &  \scriptsize  \starE    \\
\hline
\hline
\normalsize
  $P$ (days)              		& $434^{+3}_{-3}$        	& $1393^{+3}_{-3}$         		& $193.2^{+0.4}_{-0.4}$  	 	& $514^{+4}_{-4}$        \\
  $T_0$ (MJD)             	& $53955^{+23}_{-27}$    	& $54982^{+4}_{-4}$        		& $54829^{+15}_{-18}$    	    	& $53620^{+30}_{-40}$    \\
  $K$ (\!\ms)             		& $370^{+12}_{-9}$       	& $690^{+4}_{-4}$          		& $209^{+2}_{-2}$        	   	& $33^{+3}_{-3}$         \\
  $e$                     		& $0.23^{+0.09}_{-0.06}$ 	& $0.47^{+0.01}_{0.01}$    	& $0.08^{+0.05}_{-0.03}$ 	 	& $0.41^{+0.24}_{-0.12}$ \\
  $\omega$ (deg)          	& $265^{+13}_{-16}$      	& $351.3^{+0.7}_{-0.7}$    	& $196^{+30}_{-34}$      	      	& $149^{+24}_{-31}$      \\
  $m_2\sin i$ (\!\Mjup)   	& $13.9 \pm 2.4$         	& $34.1 \pm 1.1$           		& $6.1 \pm 0.7$          	        & $1.14 \pm 0.31$        \\
  $a$ (\!\au)             		& $1.14 \pm 0.08$        	& $2.47 \pm 0.03$          		& $0.66 \pm 0.04$        	       	& $1.24 \pm 0.05$        \\
  $V_0$ (\!\ms)           	& $-67263^{+17}_{-14}$   	& $-47210.6^{+2.1}_{-2.2}$ 	& $-52833.9^{+1}_{-1}$   	     	& $-36624^{+3}_{-3}$     \\
  offset (\!\ms)          		& $67163^{+30}_{-32}$    	& $47105.6^{+6.2}_{-6.2}$  	& $52897^{+11}_{-12}$    	    	& $36630^{+8}_{-8}$      \\ 
  \sjit (\!\ms)           		& 64                     		& 10.3                     			& 22                     		        & 22                     \\
  $\sqrt{\chi_\nu^2}$     	& 1.13                   		& 1.33                     			& 1.26                   		        & 1.23                   \\
  RMS (\!\ms)             	& 66                     		& 13.1                     			& 26.1                   		        & 27                     \\
  $N_{\textrm{obs}}$      	& 35                     		& 29                       			& 29                     		        & 76                     \\
\hline
\end{tabular}
\renewcommand{\arraystretch}{1}